\documentclass[twocolumn,journal]{IEEEtran}
\usepackage{color}
\usepackage{soul}
\usepackage[T1]{fontenc}
\usepackage[latin9]{inputenc}
\usepackage{array}
\usepackage{verbatim}
\usepackage{float}
\usepackage{url}
\usepackage{multirow}
\usepackage{graphicx}
\usepackage[unicode=true,
 bookmarks=true,bookmarksnumbered=true,bookmarksopen=true,bookmarksopenlevel=1,
 breaklinks=false,pdfborder={0 0 0},pdfborderstyle={},backref=false,colorlinks=false]
 {hyperref}
\hypersetup{pdftitle={Your Title},
 pdfauthor={Your Name},
 pdfpagelayout=OneColumn, pdfnewwindow=true, pdfstartview=XYZ, plainpages=false}

\makeatletter

\providecommand{\tabularnewline}{\\}
\floatstyle{ruled}
\newfloat{algorithm}{tbp}{loa}
\providecommand{\algorithmname}{Algorithm}
\floatname{algorithm}{\protect\algorithmname}

\let\oldforeign@language\foreign@language
\DeclareRobustCommand{\foreign@language}[1]{%
  \lowercase{\oldforeign@language{#1}}}

\usepackage[caption=false,font=footnotesize]{subfig}
\usepackage{algorithmic}

\usepackage{hyperref}

\usepackage{color}

\@ifundefined{showcaptionsetup}{}{%
 \PassOptionsToPackage{caption=false}{subfig}}
\usepackage{subfig}
\makeatother

\begin{document}
\title{EvoZip: Efficient Compression of Large Collections of Evolutionary
Trees}
\author{Balanand Jha, David Fern\'andez-Baca, Akshay Deepak, and Kumar Abhishek\thanks{Balanand Jha (corresponding author), Akshay Deepak, and Kumar Abhishek
are with the Department of Computer Science \& Engineering, National
Institute of Technology Patna, India, e-mail: \{balanand.cse.jrf,
akshayd, kumar.abhishek\}@nitp.ac.in.}\thanks{David Fern\'andez-Baca is with the Department of Computer Science, Iowa
State University, USA, e-mail: fernande@iastate.edu.}}
\markboth{Journal of XXX}{Balanand Jha\MakeLowercase{\emph{et al.}}: EvoZip}
\maketitle
\begin{abstract}
Phylogenetic trees represent evolutionary relationships among sets of
organisms. Popular phylogenetic reconstruction approaches typically
yield hundreds to thousands of trees on a common leafset. Storing
and sharing such large collection of trees requires considerable 
amount of space and bandwidth.  Furthermore, the huge size of 
phylogenetic tree databases can make search 
and retrieval operations time-consuming.
Phylogenetic compression techniques are specialized
compression techniques that exploit redundant topological information
to achieve better compression of phylogenetic trees. %
\begin{comment}
TASPI algorithm was the first attempt at compression of phylogenetic
trees and TreeZip is current state-of-the-art method for phylogenetic
compression.
\end{comment}
{} Here, we present EvoZip, a new approach for phylogenetic
 tree compression. On average, EvoZip achieves 71.6\% better compression
and takes 80.71\% less compression time and 60.47\% less decompression
time than TreeZip, the current state-of-the-art algorithm
for phylogenetic tree compression. While EvoZip is based on TreeZip,
it betters TreeZip due to (a) an improved bipartition and support
list encoding scheme, (b) use of Deflate compression algorithm, and
(c) use of an efficient tree reconstruction algorithm. 
EvoZip is freely available online for use by the scientific
community.
\end{abstract}


\IEEEpeerreviewmaketitle{}

\section{Introduction\label{sec:Introduction}}

\IEEEPARstart{A}{}phylogeny (or phylogenetic tree or evolutionary
tree) depicts the evolutionary relationships among a set of species
\cite{baum2008reading,deepak2014evominer}. The study of evolutionary
history of organisms has long played an important role in various
scientific research areas like comparative genetics \cite{Soltis2003},
drug discovery \cite{warren2011drug}, systematics \cite{de1988phylogenetic},
forensic analysis \cite{cracraft2005evolutionary}, and species conservation
\cite{tsang2016roles}. Phylogenies are especially useful for identifying the fast evolving lineages of
microorganisms, and tracing back
the origin of a new microorganism helps formulate the ways to handle
a new species based on the ways to handle its ancestors \cite{suzan2015metacommunity,allard2016practical,hoffmann2015tracing}.
Estimating species turnover, which has been primarily responsible
for a paradigm shift in a community ecology through prehistoric times,
is another important application of phylogenetic analysis \cite{10.2307/2400768}.

Most of the popular methods for phylogenetic reconstruction, e.g.,
maximum likelihood \cite{felsenstein1981evolutionary} and Bayesian
inference based methods \cite{huelsenbeck2001mrbayes,ronquist2012mrbayes},
produce hundreds and thousands of trees representing plausible variants
of phylogenies for a common set of taxa. Storing, mining, and sharing
all these trees often requires large amount of space and time. For
example, quite a few databases exist to store such large collections
of phylogenetic trees running into gigabytes and terabytes \cite{stBase2015,li2006treefam,huerta2013phylomedb}.
Storage and retrieval on such databases is time and space consuming.
All this points towards the need of phylogenetic tree compression
that is both time and space efficient. 

While general purpose compression techniques can be applied for phylogenetic 
tree compression, the techniques fail to exploit the semantics
of phylogenies, which limits their performance. TASPI \cite{boyer2005compressed}
appears to be the first attempt at phylogenetic compression. It works
by finding common subtrees among the set of input trees, replacing
all repeated subtrees with a \emph{reference} to its first occurrence.
TreeZip \cite{matthews2010novel} was a breakthrough in the field, greatly enhancing
the compression ratio achieved by previous algorithms. It is the current state-of-the-art phylogenetic tree compression method.  
TreeZip works by finding common bipartitions in the input tree
set and storing these bipartitions in a hash table. Only one bipartition
is stored and its all other occurrences in all of the trees are replaced
by a \emph{reference} to the stored one. This hash table and the related
data are encoded in the output file, which is much smaller 
than the original input file. The authors reported phylogenetic compression
up to 2\% \cite{matthews2010novel,matthews2015heterogeneous,matthews2010treezip}.

Here we introduce EvoZip, a new approach for phylogenetic compression. 
EvoZip is based on TreeZip, but achieves better compression
(up to 0.41\%, an improvement of 85.45\% over TreeZip) and is faster 
(an average improvement of 80.77\% in compression time and 60.47\%
in decompression time) than TreeZip. While smaller compressed
size appeals naturally, faster compression/decompression is also very
important in real time scenarios such as on-the-fly query and retrieval
operations in phylogenetic databases \cite{stBase2015,piel2000treebase,nakhleh2003requirements},
where in response to the user queries, the stored trees need to be
retrieved quickly. The improvement over TreeZip is primarily due
to modifications in the encoding scheme, use of the DEFLATE compression algorithm
\cite{oswal2016deflate,deutsch1996deflate}, and a faster tree reconstruction
algorithm (used during the decompression stage). Further, EvoZip also
supports internal node labels in Newick trees, a feature that is not supported
by TreeZip.

The rest of the paper is organized as follows. In Sec.\ \ref{sec:PhyloTreesNewick}, we review phylogenetic trees and their representation using the Newick format. In Sec. \ref{sec:The-TreeZip-Algorithm}
we present a detailed description of the TreeZip algorithm. Section
\ref{sec:Our-Contribution} describes EvoZip algorithm highlighting
its improvements over TreeZip. In Sec. \ref{sec:Results-and-discussion}
we compare EvoZip and TreeZip with respect to the compressed file size and compression/decompression time. Lastly
in Sec. \ref{sec:Conclusions-and-future} we conclude the paper and
discuss the scope of future work.

\begin{comment}
As suggested by the referenced work and our observation, semantic
compression offers significant improvements over general purpose compression
mechanisms in terms of compression ratio for phylogenetic trees.
\end{comment}

\section{Phylogenetic Trees and the Newick Format\label{sec:PhyloTreesNewick}}

\begin{figure}[tbh]
\centering{}\subfloat[\label{fig:SimpleUnweightedTree}Unweighted]{\includegraphics[scale=0.6]{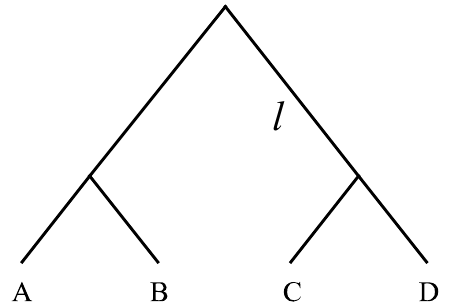}

}\ \ \ \ \ \ \subfloat[\label{fig:SimpleWeightedTree}Weighted]{\includegraphics[scale=0.6]{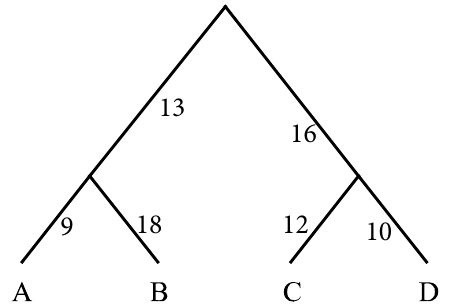}}\ \ \subfloat[\label{fig:Simple-unweighted-with-internal}Unweighted with internal
node labels]{\includegraphics[scale=0.6]{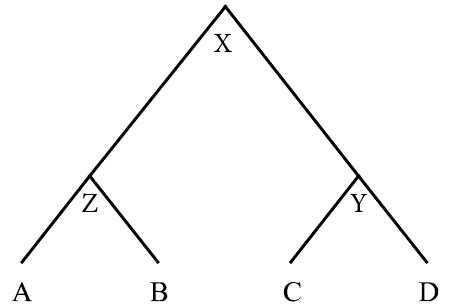}}\caption{\label{fig:Simple-tree-ex}Simple unweighted and weighted evolutionary
trees}
\end{figure}

An evolutionary tree is a leaf-labeled rooted tree (see Fig. \ref{fig:Simple-tree-ex})
where terminal nodes (leaves) represent taxa of interest and internal
nodes represent ancestors. Evolutionary relationship are indicated
by placing ancestors closer to the root node than any of their descendants.
For example, in Fig. \ref{fig:Simple-unweighted-with-internal}, $Y$
is an ancestor to both $C$ and $D$, while $X$ is an ancestor to
all the three, i.e., $Y$, $C$, and $D$. Often the ancestors are
not identified (as in Fig. \ref{fig:SimpleUnweightedTree} and \ref{fig:SimpleWeightedTree})
due to a lack of information or interest. When identified, the tree
has internal node labels too (as in Fig. \ref{fig:Simple-unweighted-with-internal}).
Though commonly found as unweighted binary rooted trees, an evolutionary
tree can be binary or n-ary, rooted or unrooted, and weighted (Fig.
\ref{fig:SimpleWeightedTree}) or unweighted (Fig. \ref{fig:SimpleUnweightedTree})
\cite{baum2008reading}. In all their variations, evolutionary trees
are always unordered trees; i.e., the order of children (in rooted
trees) or neighbors (in unrooted trees) does not matter.  In this paper, a phylogenetic tree refers
to a rooted phylogenetic tree, unless stated otherwise.

Every edge of a phylogenetic tree partitions the set of taxa into
two disjoint subsets. For example, in Fig. \ref{fig:SimpleUnweightedTree},
the edge labeled $l$ partitions the set of taxa into $\{A,B\}$ and
$\{C,D\}$. This \emph{bipartition }induced by edge $l$ is represented
as $AB|CD$. The set of all bipartitions uniquely identifies a phylogenetic
tree \cite{matthews2010novel}. For example, the tree shown in Fig.
\ref{fig:Simple-tree-ex} is uniquely identified by its bipartitions
$AB|CD,$ $A|BCD,$ $B|ACD,$ $C|ABD,$ $D|ABC$. The last four bipartitions
have a partition with a single leaf and are called \textsl{``trivial'',}
because such bipartitions are present in every possible tree on the
given set of taxa. For a given phylogenetic tree, either of the subsets
of a bipartition uniquely identifies the bipartition. For example,
either of $\{A,B\}$ and $\{C,D\}$ uniquely identifies the bipartition
$AB|CD$. Hence, we represent a bipartition by just one of its subsets.
To ensure uniqueness in the representation of a bipartition, we always
choose the set of leaf nodes that are descendants of the bipartition
edge. For example, in Fig. \ref{fig:SimpleUnweightedTree}, for the
bipartition $AB|CD$ induced by edge $l$, we choose $CD$ to represent
the bipartition. We define the \emph{cardinality }of a bipartition
as the number of leaf nodes that are descendants of the bipartition
edge. For example, in Fig. \ref{fig:SimpleUnweightedTree}, for the
bipartition $AB|CD$ induced by edge $l$, and represented by $CD$,
its cardinality is same as the cardinatily of set $\{C,D\}$, i.e.,
two.

EvoiZip, our proposed compression algorithm, takes as input  phylogenetic trees 
represented using the popular Newick format\footnote{\url{http://evolution.genetics.washington.edu/phylip/newicktree.html}}.  
%While Newick represents rooted trees, unrooted trees can also
%be stored in Newick format by introducing a root node. The root node
%is removed later at the time of need to retrieve the original unrooted
%tree. 
The Newick format represents phylogenetic trees as strings, 
using parenthesis to delimit nodes, and commas
to separate the children of a node. For example, the Newick representation
of the tree shown in Fig. \ref{fig:SimpleUnweightedTree} is $((A,B),(C,D));$,
where `;' marks the end of the tree. Newick format can also represent
weighted phylogenetic trees, where the weights \textemdash{} represented
by real numbers \textemdash{} are placed just after a node separated
by a colon. For example, the tree in Fig. \ref{fig:SimpleWeightedTree}
is a weighted tree. The corresponding Newick representation is $((A:9,B:18):13,(D:10,C:12):16):10;$.
Apart from the terminal nodes, the internal nodes can also be labeled,
which account for the name of a known ancestor. For example, for the
tree in Fig. \ref{fig:Simple-unweighted-with-internal}, its Newick
representation is $((D,C)Y,(B,A)Z)X;$. Since, all trees stored in
Newick format are rooted and our work is about compressing trees in
Newick format,

\begin{comment}
A file containing trees in newick format is a simple text file. General
purpose compression tools like 7Zip \cite{pavlov20077zip}  compress
these files like they do to any other text file, i.e., by minimizing
redundant symbols in the file. However, as pointed out in \cite{matthews2010novel},
general purpose compression algorithms do not recognize the evolutionary
relationship in the trees (e.g., the bipartition information). TreeZip
is a phylogenetic compression algorithm that performs \emph{semantic
compression} of phylogenetic trees, i.e., it performs compression
by removing redundant bipartition information in a collection of trees
on a common leaf set. It also employs additional algorithmic techniques
from data compression to achieve further compression. Motivated by
this, the semantic compression also lies at the heart of our proposed
work.
\end{comment}

\section{The TreeZip Algorithm\label{sec:The-TreeZip-Algorithm}}

TreeZip \cite{matthews2010novel,matthews2015heterogeneous,matthews2010treezip},
is the current state-of-the-art algorithm for phylogenetic compression, 
works by exploiting redundancy among the bipartitions in large collections of phylogenetic
trees on a common leafset. The key idea in TreeZip is
to organize each bipartition from a collection of phylogenetic trees
in a universal hash table, where each entry of a bipartition in the
hash table also stores the list of trees containing that bipartition.
Each of the input trees is parsed and the corresponding bipartitions
are simultaneously inserted into the hash table. The compressed output
file contains dataset properties such as the number of taxa, taxon
labels (names of taxa or leaf nodes), and the number of trees, followed
by bipartitions and their \emph{support lists} (the list of trees
containing the bipartition). The bipartitions and their support lists
are encoded using a combination of run-length and delta encoding schemes.
%This achieves tremendous compression with respect to its original
%size.
Next, we review TreeZip's randomized universal hash function, along with its algorithm to create
and populate the hash table, and TreeZip's encoding schemes for bipartitions and support lists.

\subsection{TreeZip's hash function}

TreeZip adapts the idea of hashing bipartitions from Amenta et al.'s
\cite{amenta2003linear} randomized linear-time algorithm for generating a majority rule
tree. We note that the authors of TreeZip have also used this
idea in HashCS \cite{sul2009experimental} ---  which generates the 
majority-rule tree for a collection of trees --- and HashRF
\cite{sul2008experimental} --- which constructs
the Robinson-Foulds distance matrix \cite{robinson1981comparison} for a collection of trees.

For a given tree on $n$ leaves, where an implicit ordering on the
set of leaves is assumed, let $B=(b_{1}...b_{n})$ be the \textbf{n}-bit
bitstring representation of a bipartition in the tree. Here, bit $b_{i}$
is set to $1$ if the $i^{\mathrm{th}}$ taxon is present in the bipartition
and $0$ otherwise. In this regard, we define a special \emph{root
bipartition }corresponding to the root node as the bit string representation
with every bit set to 1. The following universal hash function is
used to hash bipartitions:
\[
h\left(B\right)\mathrel{=}\mathop{\sum b_{i}a_{i} \bmod}m
\]
where
$B=(b_{1}...b_{n})$ is the \textbf{n}-bit bitstring representation
of the bipartition, 
$m$ is a large prime number, and\\
$A=(a_{1}...a_{n})$ denotes a set of random integers between $0$
and $m-1$.

Clearly, the larger is the value of $m$, the smaller are the chances
of two different bipartitions hashing to the same code. However, since $m$ 
equals the number of bins in the hash table, $m$ cannot be too large. To
address this, two such hash functions, $h_{1}$ and $h_{2}$, corresponding
to $m_{1}$ and $m_{2}$, are used to hash each of the bipartitions.
While $h_{1}$ determines the bin index in the hash table where the
corresponding bipartition is stored, $h_{2}$ is used to resolve a
collision of two different bipartitions having a common $h_{1}$ value,
i.e., indexed in the same $h_{1}^{\mathrm{th}}$ bin of the hash table.
With this, the probability of collision at the level of $h_{2}$ becomes
$1/m_{1}m_{2}$. The value of $m_{1}$ determines the size of the hash table,
and can be adjusted depending on memory considerations.
The value of $m_2$ is chosen to be a very large number, $m_{2}>>m_{1}$, 
to minimize the probability of collision.

A nice property of the above hash function is that it can be computed
%\comm{Does this refer only to $h_1$ or also to $h_2$?}
for all the bipartitions in a tree in $O(n)$ time, where $n$ is
the number of leaves. Compare this to the naive approach that will
take $O(n^{2}/w)$, where $w$ is the size of the machine word. This
is possible because given the hash codes of the children of a node,
the hash code of the parent node can be calculated in $O(1)$ time.
Let a node with bit string representation $B$ have two children with
$B_{L}$ and $B_{R}$ as their respective bitstring representations.
Then, given $h_{1}(B_{L})$ and $h_{1}(B_{R})$, $h_{1}(B)$ can be
computed in $O(1)$ time as:
\[
h_{1}(B)=\left(h_{1}(B_{L})+h_{1}(B_{R})\right)\, \bmod\,m_{1}
\]

Same applies to the second hash function $h_2$ as: 
\[
h_{2}(B)=\left(h_{2}(B_{L})+h_{2}(B_{R})\right)\, \bmod\,m_{2}
\]Once the hash table is populated, it is encoded and written to the
output file. For each bipartition, its bitstring representation and
the support list are stored in the output file. 

\subsection{Encoding bipartitions} 
\begin{figure}[tbh]
\includegraphics[scale=0.5]{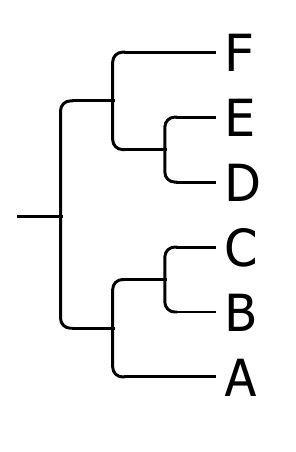}\includegraphics[scale=0.5]{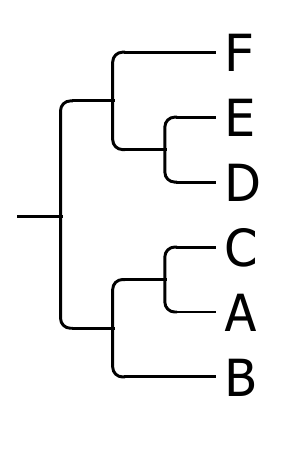}\includegraphics[scale=0.5]{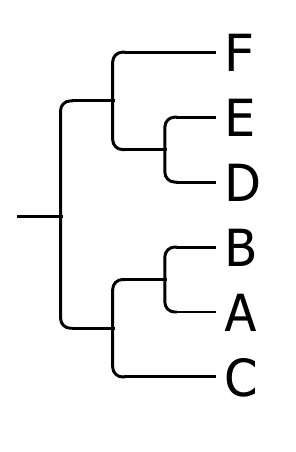}\includegraphics[scale=0.5]{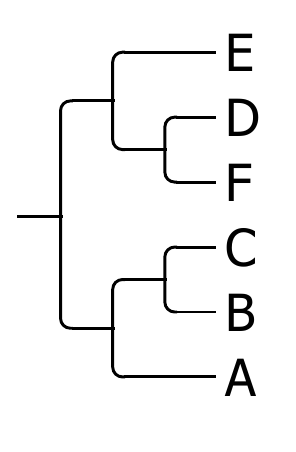}\includegraphics[scale=0.5]{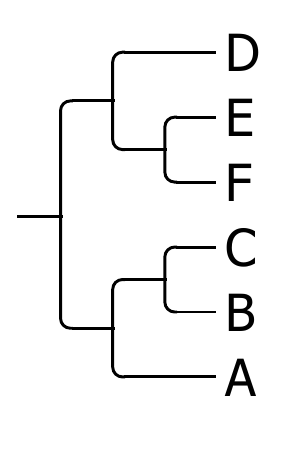}\includegraphics[scale=0.5]{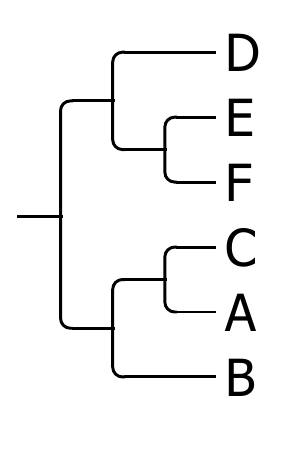}\caption{\label{fig:example-trees}A collection of six tree on a common leafset.}
\end{figure}

As discussed previously, a bipartition
is represented by the subset of leaf nodes that are descendants of
the bipartition edge. Thus, in the bit string representation,
only the bits corresponding to the leaf nodes present in this subset
are turned ON. For example, the bitstring representation of a
bipartition $ABC|DEF$ is 000111. %
\begin{comment}
First all the taxon names are arranged in a fixed order (derived from
the taxon order in the first tree) for all bipartitions. This order
is same as the order in which all the taxon names are listed in the
header part of the .trz file. A fixed length bitstring of size equal
to the number of taxa in the tree is initialized. While encoding,
those taxa which occur left to the separator in bipartition are represented
by 0 while those occurring to the right are represented by 1.
\end{comment}
Such a bitstring representation is inherently highly redundant, so
storing the bitstrings as such in the compressed file would not be
space efficient. Hence, the authors apply run length encoding where
they represented long runs of 0s and 1s as characters $L$ and $K$
respectively, followed by the respective counts of $0s$ and $1s$.
For single occurrences of $0s$ and $1s$ in the bitstring, they used
symbols $A$ and $B$ respectively. For instance, consider bipartition
$ABCF|DE$ of the second tree in Fig. \ref{fig:example-trees} having
bitstring representation $000110$. It has a continuous stretch of
three $0s$, two $1s$, and a single $0$. The first three $0s$ are
encoded as $L3$, the next two $1s$ are encoded as $K2$, and finally
the single $0$ is encoded as $A$. The complete encoded bitstring
representation of this bipartition is $L3K2A$. Table \ref{tab:List-of-bipartitions}
shows bitstring representation and run length encoding of all the
bipartitions of all the trees in the tree set shown in Fig. \ref{fig:example-trees}.

\begin{table}[tbh]
\caption{\label{tab:List-of-bipartitions}List of non-trivial bipartitions
of trees in Fig. \ref{fig:example-trees}}

\medskip{}
\begin{tabular}{|c|c|c|c|c|c|}
\hline 
Tree 0 & Bitstring & RLE & Tree 1 & Bitstring & RLE\tabularnewline
\hline 
\hline 
ABC|DEF & 000111 & L3K3 & ABC|DEF & 000111 & L3K3\tabularnewline
\hline 
BC|ADEF & 100111 & BL2K3 & AC|BDEF & 010111 & ABAK3\tabularnewline
\hline 
ABCF|DE & 000110 & L3K2A & ABCF|DE & 000110 & L3K2A\tabularnewline
\hline 
\end{tabular}\medskip{}
\begin{tabular}{|c|c|c|c|c|c|}
\hline 
Tree 2 & Bitstring & RLE & Tree 3 & Bitstring & RLE\tabularnewline
\hline 
\hline 
ABC|DEF & 000111 & L3K3 & ABC|DEF & 000111 & L3K3\tabularnewline
\hline 
AB|CDEF & 001111 & L2K4 & BC|ADEF & 100111 & BL2K3\tabularnewline
\hline 
ABCF|DE & 000110 & L3K2A & ABCE|DF & 000101 & L3BAB\tabularnewline
\hline 
\end{tabular}\medskip{}
\begin{tabular}{|c|c|c|c|c|c|}
\hline 
Tree 4 & Bitstring & RLE & Tree 5 & Bitstring & RLE\tabularnewline
\hline 
\hline 
ABC|DEF & 000111 & L3K3 & ABC|DEF & 000111 & L3K3\tabularnewline
\hline 
BC|ADEF & 100111 & BL2K3 & AC|BDEF & 010111 & ABAK3\tabularnewline
\hline 
ABCD|EF & 000011 & L4K2 & ABCD|EF & 000011 & L4K2\tabularnewline
\hline 
\end{tabular}
\end{table}

\subsection{Encoding Support Lists}
The support list of a bipartition
consists of the trees containing the bipartition. Here, each tree is represented
by its identifier, its \emph{tree id}, which is an integer. This
list of integers is stored in the ascending order in the hash table
where the corresponding bipartition is stored. 

A particular bipartition may be present in some trees and absent in others. If a bipartition
is present in majority of the trees, then storing the list of trees
where the bipartition is absent can be more space-efficient than storing its support
list. Motivated by this, Matthews et al.\ coin two terms, \emph{in-set} and
\emph{out-set}. The in-set is the set of trees containing a particular
bipartition and the out-set is its compliment. Only the smaller of the
two is stored in the output compressed file. A marker, ``+'' for
in-set and ``-'' for out-set, is written in the output file to indicate
the corresponding choice. 

To store a list of tree ids, first the
in-set/out-set marker is written followed by the first tree id. For
subsequent tree ids, only the difference with respect to the previous tree id
is written in the output file. To avoid the need of a special character
to distinguish between subsequent entries in this list, the authors
used a trick to distinguish the first digit of each entry by
using letters A to J for the digits 0 to 9. For example, for the bipartition
ABCD|EF present in trees 4 and 5 in Fig. \ref{fig:example-trees}
and in Table \ref{tab:ListBipartSupList}, the first tree id $4$
is encoded in entirety as \emph{E}. For the next tree id, 5, only its
difference with respect to the previous tree id is encoded, i.e.,
$5-4=1$ is encoded as $B$. The final encoded list is $EB$. For
each bipartition of the tree set shown in Fig. \ref{fig:example-trees},
Table \ref{tab:ListBipartSupList} shows the corresponding bitstring
representation, the run length encoded bitstring, its support list,
and the delta encoded support list.

\begin{table}[tbh]
\caption{\label{tab:ListBipartSupList}List of bipartitions and their support
list}
\medskip{}
\begin{tabular}{|c|c|c|c|c|c|}
\hline 
Bipartition & Bitstring & RLE & Tree Ids & +/- & Support List\tabularnewline
\hline 
\hline 
ABCD|EF & 000011 & L4K2 & 4,5 & + & EB\tabularnewline
\hline 
ABCE|DF & 000101 & L3BAB & 3 & + & D\tabularnewline
\hline 
ABCF|DE & 000110 & L3K2A & 0,1,2 & + & ABB\tabularnewline
\hline 
ABC|DEF & 000111 & L3K3 & 0,1,2,3,4,5 & - & 0\tabularnewline
\hline 
AB|CDEF & 001111 & L2K4 & 2 & + & C\tabularnewline
\hline 
AC|BDEF & 010111 & ABAK3 & 1,5 & + & BE\tabularnewline
\hline 
BC|ADEF & 100111 & BL2K3 & 0,3,4 & + & ADB\tabularnewline
\hline 
\end{tabular}
\end{table}

The Newick representation of the input trees and the contents of the
final compressed file are shown in Table \ref{tab:Contents-of-the-trz-file}.
The first line of the output file contains the list of all the taxa
in the input tree set. The second line stores the number of trees.
The third line shows the total number of bipartitions in the entire
tree set. The subsequent lines represent encoded bipartitions and
the corresponding support list; each such line consists of four fields:
the encoded bipartition, in-set/out-set marker, number of entries
in the support list, and finally, the encoded support list.

\begin{table}[tbh]
\caption{\label{tab:Contents-of-the-trz-file}Contents of the compressed file}

\begin{tabular}{|>{\raggedright}m{0.4\columnwidth}|>{\raggedright}m{0.4\columnwidth}|}
\hline 
(A,(B,C),(D,E),F);\\
(B,(A,C),(D,E),F);\\
(C,(A,B),(D,E),F);\\
(A,(B,C),(D,E),F);\\
(A,(B,C),(E,F),D);\\
(B,(A,C),(E,F),D); & 1. TAXA A:B:C:D:E:F\\
2. NTREES 6\\
3. NBIPARTITIONS 7\\
4. L4K2 +:2:EB\\
5. L3BAB +:1:D\\
6. L3K2A +:3:ABB\\
7. L3K3 -:0:\\
8. L2K4 +:1:C\\
9. ABAK3 +:2:BE\\
10. BL2K3 +:3:ADB\tabularnewline
\hline 
\end{tabular}
\end{table}

\section{The EvoZip Algorithm\label{sec:Our-Contribution}}

EvoZip is based on TreeZip.  EvoZip outperforms TreeZip, in terms of both 
compressed size and compression/decompression time, using the following ideas.
% \textendash{} the state-of-the-art algorithm for phylogenetic compression. 
%EvoZip improves upon TreeZip with respect to the following:
\begin{itemize}
\item \textbf{Improved encoding scheme of bipartitions}: The bipartition
encoding scheme of EvoZip improves the encoding scheme of TreeZip
by better utilization of character space by employing single byte
values as markers for different categories of byte values during compression.
\item \textbf{Improved encoding scheme of support lists}: EvoZip utilizes
fixed-size delta encoding technique to compress the list of trees
that contain a particular bipartition.
\item \textbf{Use of Deflate compression algorithm}: EvoZip further compresses
the encoded data using the Deflate compression algorithm \cite{oswal2016deflate}, 
which combines LZ77 \cite{ziv1977universal} and Huffman encoding
\cite{huffman1952method}.
%it first encodes the data using LZ77, and
%then using Huffman encoding.
\item \textbf{Efficient tree reconstruction}: EvoZip improves on TreeZip's 
decompression method by using Gusfield's \cite{gusfield1991efficient}
algorithm for reconstructing phylogenetic trees from bipartitions
during decompression. 
%To the best of our knowledge, Gusfield's algorithm
%is the fastest approach for reconstructing a phylogenetic tree from
%its constituent bipartitions.
\end{itemize}
Using the above ideas, EvoZip achieves average improvements, relative to TreeZip,
of 71.60\% on the  compressed size
of the output file and of 80.77\%
and 60.47\% on the compression and decompression times, respectively.
In the rest of this section, we discuss in detail the techniques used by EvoZip.

\subsection{Improved bipartition encoding scheme}

TreeZip's encoding scheme was explained in Sec.\
\ref{sec:The-TreeZip-Algorithm}. For example, TreeZip encodes `0001101'
as `L3K2AB'. EvoZip improves upon the bipartition encoding of TreeZip
in two ways. First, TreeZip encodes the count of consecutive
zeroes (or ones) as a base 10 number, where each digit of the number
is stored as a character. For the example mentioned, the `000' part
is encoded as character string `L3', where `L' is the marker character
for zeros. This prevents the full utilization of a character space.
EvoZip overcomes this limitation by encoding the count of consecutive
zeros (or ones) as a base 2 number, where each digit of the number
is encoded as a bit. This allows better utilization of a character
space. For example, to encode 120, TreeZip uses three bytes whereas
EvoZip takes only one byte, i.e., $120_{10}=1111000_{2}$, where each
digit of 1111000 is encoded as a bit. Second, TreeZip's run-length encoding
of a bipartition becomes particularly inefficient for long
sequences of non-consecutive zeros and ones. TreeZip encodes a single
occurrence of zero (one) as the character `A' (`B'). Hence, TreeZip
encodes `10101010' as string `ABABABAB'. Again, this prevents the
full utilization of the corresponding character space because to indicate
a 0/1, one bit of space is sufficient. EvoZip overcomes this limitation
by encoding sequences of non-consecutive zeros and ones as 8-bit sequences,
where each character of the bit-string representation is represented
by a bit. For example, TreeZip encodes `10101010' as string `ABABABAB',
taking eight bytes, whereas EvoZip encodes each digit of `10101010'
as a single bit taking only one byte.

\begin{figure}[tbh]
\centering{}\includegraphics[scale=0.35]{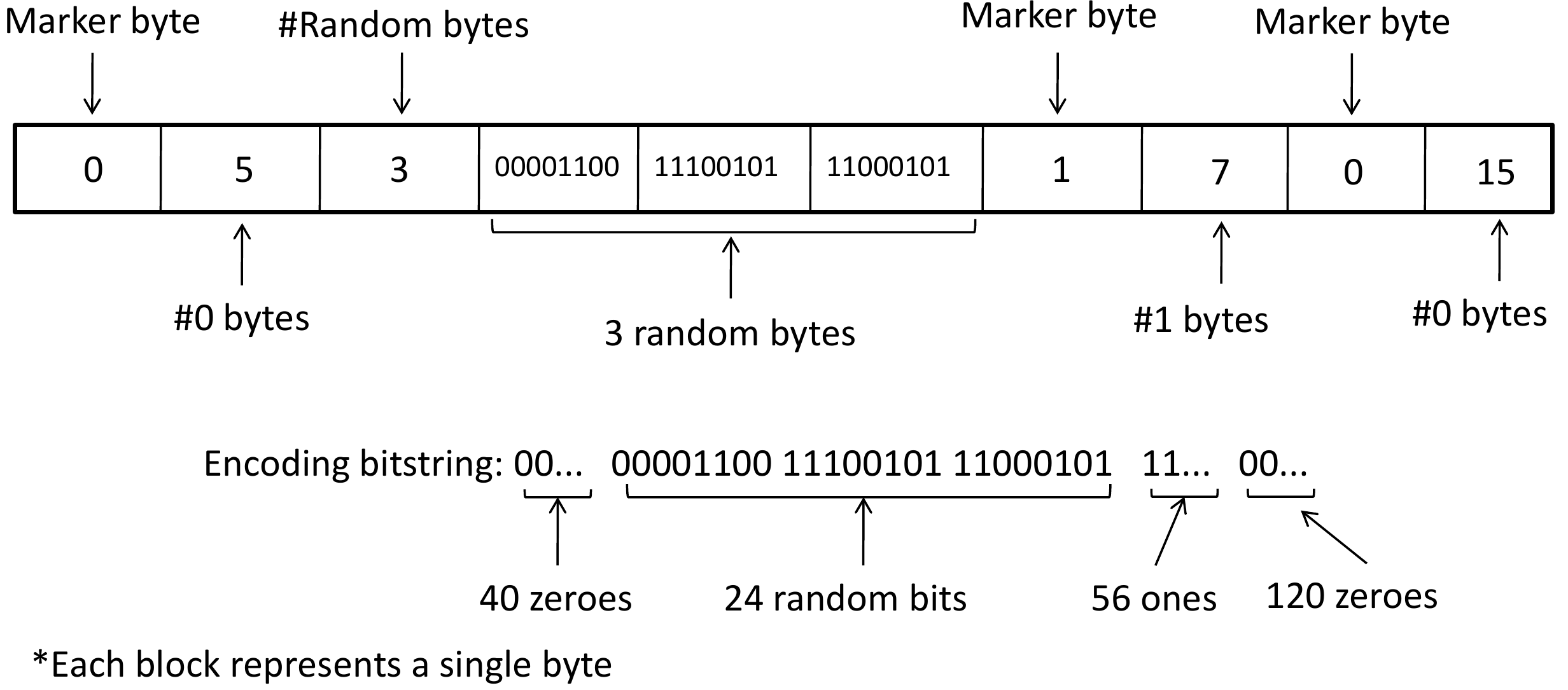}\caption{\label{fig:Bipartition-encoding-scheme}Bipartition encoding scheme.}
\end{figure}

Figure \ref{fig:Bipartition-encoding-scheme} explains the bipartition
encoding scheme of EvoZip with an example. There are three possibilities
of consecutive 0s and 1s in a bipartition bitstring. There may be
a long sequence of 0s, 1s, or a random stretch of non-consecutive
sequence of 0s/1s. We indicate this using a single byte number as
a marker. A $0$ value indicates that the number in the next byte
is the number of chunks of contiguous $0s$ in the bitstring. A $1$
value indicates the same for $1s$. Any other value indicates the
number of 8-bit sequences of non-consecutive 0s/1s; these non-consecutive
sequences are written just after. For simplicity, a bipartition bitstring
is processed in chunks of 8-bit sequences. In this regard, only an
8-bit chunk of all 1s/0s is considered consecutive. For example, consider
the bitstring shown in Fig. \ref{fig:Bipartition-encoding-scheme}.
Reading 8-bit chunks from left to right, we first encounter a subsequence
of 40 zeros. This is encoded as `05' taking two bytes. Here, `0' indicates
that the subsequence being encoded consists of 8-bit sequences of
consecutive 0s and `5' indicates the number of consecutive 8-bit sequences
of 0s. Next, we encounter a random stretch consisting of three 8-bit
sequences of non-consecutive sequence of 0s/1s. To encode this, we
set the marker as `3' and write the subsequent random bits just after
this marker. Next, we have 7 chunks of 1s and 15 chunks of 0s that
are written just after following the ``marker \textendash{} count''
pattern. EvoZip requires ten bytes to store this encoded bipartition.
TreeZip encodes this bipartition as L44K2L2K3L2BAK3L3BAK57A120 (see
Sec. \ref{sec:The-TreeZip-Algorithm}), requiring a total of 26 bytes.
EvoZip has a gain of 16 bytes.

\subsection{Improved support list encoding scheme}

The \textit{support list} of a bipartition is the list of trees having
that bipartition. Each tree is represented by its unique integral
identifier called \emph{tree id}. Both TreeZip and EvoZip
employ \emph{delta encoding} \cite{mogul2001delta}.  
The first tree id in a support list is stored as is. For each 
subsequent tree id, only its difference --- i.e., its
\emph{delta}) --- with respect to the previous tree id is stored.  
TreeZip encodes the difference as a base 10 number, where each digit
of the number is stored as a character.  In contrast, EvoZip encodes it as a base
2 number, where each digit of the number is encoded as a bit. This
results in a better utilization of space as was done while encoding
bipartitions. First, EvoZip tries to encode the difference in a single
byte. Only if the delta cannot be accommodated in a byte, i.e., if
it exceeds the value of 255, EvoZip uses more bytes. This saves space
when compared with a scheme that reserves a fixed number of bytes
for storing each delta value. 

\begin{figure}[tbh]
\centering{}\includegraphics[scale=0.35]{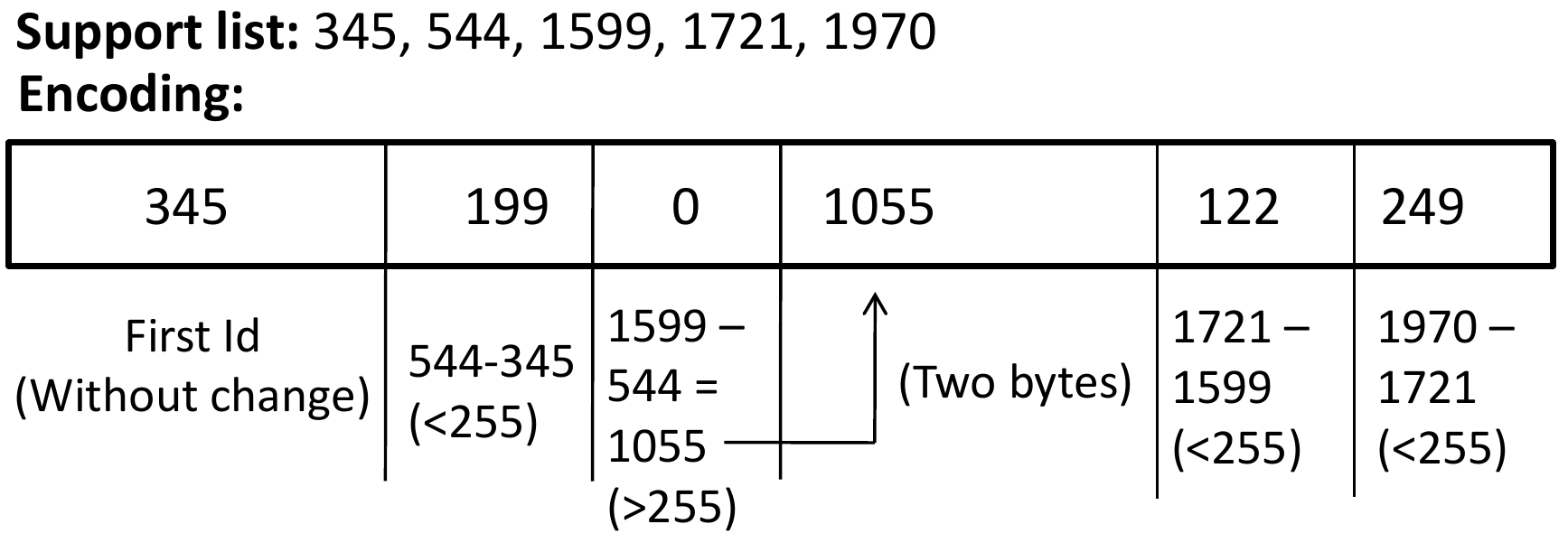}\caption{\label{fig:Support-list-encoding}Support list encoding scheme.}
\end{figure}

For example, consider the support
list shown in Fig. \ref{fig:Support-list-encoding}. The first tree
id (345) fits in the first two bytes and is stored as such. For the
next tree id 544, having a difference of 199 from previous tree id,
its difference, i.e., 199, is stored in a single byte. For the next
tree id 1599, with a difference of 1055 from the previous tree id,
its difference, i.e., 1055, exceeds the capacity of a single byte
(255). Hence, EvoZip sets the next byte to zero to indicate this event
and stores the difference in the next two bytes. Further, the next
two tree ids with a difference of 122 and 249 are stored in a single
byte each. EvoZip takes a total of eight bytes to store this support
list. TreeZip encodes this support list as D45B99B055B22C49 
(see section \ref{sec:The-TreeZip-Algorithm}), which requires 16 bytes.
In comparison, EvoZip achieves a gain of 8 bytes for the given
example. This gain becomes increasingly larger as the number 
of trees and bipartitions increases. 
%Fig.\ \ref{fig:Support-list-encoding} illustrates EvoZip's support list encoding scheme.

\subsection{The Deflate algorithm}

\begin{figure}[tbh]
\includegraphics[scale=0.42]{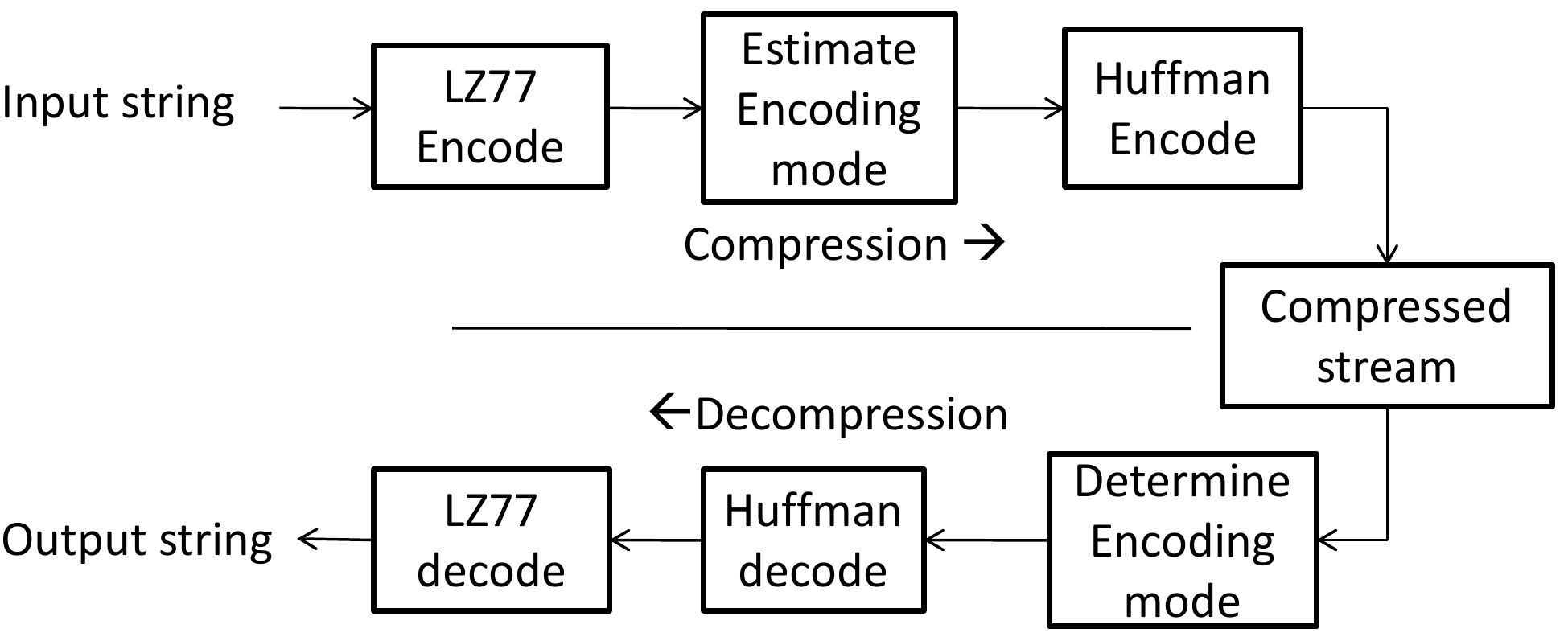}\caption{\label{fig:Block-diagram-of-deflate}Block diagram of deflate algorithm}
\end{figure}

The encoded bipartitions and support lists are written in an intermediate
buffer. Though the space used by this buffer is much smaller than the
space required to store the trees in the Newick format, it is further
compressed with the use of the Deflate algorithm \cite{oswal2016deflate}.
Deflate
% is possibly one of the best examples of perfect cooperation
%between two different algorithms. It 
works by first compressing the
input stream using the LZ77 \cite{ziv1977universal} algorithm and then
encoding this compressed stream using Huffman encoding \cite{huffman1952method}.
%LZ77 is a widely known dictionary based compression algorithm that
%searches for redundant string in a fixed length of input stream. Huffman
%coding is a prefix code that represents symbols from a fixed set of
%alphabets in the form of bit sequence. The symbols of the alphabet
%are placed at the leaf nodes of a binary tree and edges are given
%binary zero or one value. The bitstring code of a symbol is calculated
%by traversing the tree from the root to the leaf node of interest
%and placing the binary value of each intermediate edge just right
%to the previous edge value. 
A block diagram of the Deflate algorithm
is shown in Fig. \ref{fig:Block-diagram-of-deflate}. We now briefly review the relevant aspects of LZ77 and 
Huffman encoding.
%\comm{Do we need such a thorough review of LZ77 and Huffman encoding?}

\subsubsection{LZ77 algorithm}

The well-known LZ77 string compression algorithm  \cite{ziv1977universal} 
minimizes redundancy in the input stream by identifying
repeatedly occurring substrings in the input stream, retaining
only the first occurrence, and replacing all the other occurrences
by a distance-length vector. This vector points to the start of the
occurrence of a matched string by storing backward distance of the
first symbol of the string from the current position. The length component
of this vector tells how long is the matching string. %The complexity
%of backward search limits the distance to search for the matching
%string and its length in the input stream. In order to keep the compression
%time within practical limits, 
A \textit{sliding
window} of fixed length limits the backward search distance (typically 32KB) for finding a match. The window
is divided into two parts: (1) search buffer or dictionary and (2)
look ahead buffer. The search buffer contains the part that has already
been processed. New data to be searched is loaded into the look ahead
buffer, which is compared with the data in the search buffer to find
a match. 
%Algorithm \ref{alg:LZ77-encoding} describes the LZ77 encoding process.
In the Deflate algorithm, backward-distance and length are limited to 32K bytes and 258 bytes respectively \cite{deutsch1996deflate}.

\subsubsection{Huffman encoding}

Huffman encoding is a well known prefix based code for compressing
bitstreams. The peculiarity of this algorithm is that it assigns variable
length codes depending on the frequency of symbols in the input stream.
Highly frequent symbols get smaller codes while those occurring less
frequently get longer codes. It is called prefix based code because
each symbol gets its own prefix code that is unique to that symbol
only in that particular encoding instance, and no code is a prefix
of another code. %For example, if symbol `a' is represented by code
%010, no other symbol will get the code 01 as it is a prefix of the
%code for symbol `a' and including this in the code alphabet will make
%the code ambiguous. It will be impossible to decode a bitstring like
%0100101010 if such ambiguity is allowed in the code because the decoder
%will not be able to determine whether the starting symbol in the code
%is `a' or `b' (if code 01 denotes symbol `b'). 
The unique prefix nature
of this code avoids the need of end of the character markers in the
encoded string. The output stream from LZ77 is fed to the Huffman
encoder. In order to determine a unique code for each character in
the file, the algorithm first gets frequencies of each character and
then constructs a Huffman tree. A Huffman tree is a binary tree having
each character from the text as a leaf node. All the edges in the
tree are labelled with either ``0'' or ``1''. If a node has two
descendant nodes, then the edge connecting the current node to its
left descendant will be labelled as ``0'' and that of right descendant
as ``1''.

%To create a Huffman tree, all the symbols are considered to be
%leaves of a binary tree. These leaves are sorted by increasing
%order of their frequencies or code-lengths. The first two leaves having
%minimum frequencies are selected and a new internal node is created.
%The two leaf nodes are made children of the newly formed node; the
%leaf with lower frequency being left child and the other as the right
%child. Again the nodes are sorted and a new internal node is created
%from two nodes with minimum frequencies. When all the leaves are processed,
%we get the desired Huffman tree. 
The bitstring representation of a
character can be obtained from this tree by traversing the tree from
the root node to the leaf of interest along the shortest path and
appending the label of each edge in the path. Using this tree, the
bitstring code for each symbol can be generated. %An example of coding

\subsubsection*{Combining LZ77 and Huffman in `Deflate'}

The output of LZ77 algorithm is fed to the Huffman encoder, which
breaks it into smaller blocks. Each of these blocks is encoded independently.
The Huffman encoder in Deflate employs three modes of compression.
Each block starts with a three bit header to indicate the mode of
compression of the current block. The first bit indicates whether
the current block is the last block of the data or not. It is set
to true only for the last block of the data. The values of the next
two bits denote the mode of compression used. The value 00 of these
two bits indicate that the block contains incompressible data. The
value 01 indicates that the block is compressed using a common Huffman
code, which is available at the decoder end itself (also known as
``fixed Huffman code''). Hence, the Huffman code table, which is
required for reconstructing the Huffman code tree, need not be present
in the encoded block. The value `10' indicates that the block is compressed
with Huffman code with dedicated tree for each block (also known as
``dynamic Huffman code''). The Huffman code table must be sent in
each block. A value of `11' is reserved for unexpected situations
(like errors) and is not used for encoding/decoding.

As discussed before, the LZ77 encoded stream consists of distance-length-literal
triplets. The length and literal segments are encoded into a single
Huffman tree while the distance is encoded in another Huffman tree.
Each block of the compressed Huffman code stream contains three header
bits followed by a pair of Huffman tables and the compressed stream.
The size of a block is not fixed for compressed blocks (the blocks
in which the block type indicator bits are anything other than 00),
however, those blocks containing uncompressed data (indicator bits
00) are limited to 65,535 bytes. Once the limit has been reached,
the current block must be terminated and a new block must begin.

\subsubsection{Tree reconstruction algorithm}

The tree reconstruction algorithm used by EvoZip is based on Gusfield's
\cite{gusfield1991efficient} $O(nm)$ time tree reconstruction algorithm,
where $n$ is the number of taxa and $m$ is the number of bipartitions.
To the best of our knowledge, Gusfield's algorithm is the fastest
algorithm for reconstructing trees from its constituent bipartitions.
It is based on the following observation for deriving parent-descendant
relationship among a set of bipartitions: if two bipartitions have
a common leaf descendant (\emph{overlapping bipartitions})\emph{,
}the one with the higher cardinality is the parent of the other\emph{.}
Algorithm \ref{alg:Tree-reconstruction-algorithm} describes the reconstruction
algorithm. First the set of $\{$non-trivial bipartitions $\cup$
root bipartition $\}$ is sorted by ascending order
of the cardinalities of bipartitions, the root bipartition appearing
last in the sorted list (lines \ref{reconst.1}--\ref{reconst.2}).
The tree is reconstructed iteratively in terms of paths from leaf
nodes to the root, each such a path being reconstructed in an iteration
(lines \ref{reconst.forStart}--\ref{reconst.forEnd}). Two or more
of such paths can overlap. While reconstructing a path, only the non-overlapping
portion of a path that has not been constructed before, is constructed
(lines \ref{reconst.elseStart}--\ref{reconst.elseEnd}). The construction
of a path is stopped the moment an already constructed node \textendash{}
belonging to the overlapping part \textendash{} is encountered in
the leaf-to-root path (lines \ref{reconst.ifStart}--\ref{reconst.ifStop}).

\begin{figure}[tbh]

\includegraphics[scale=0.4]{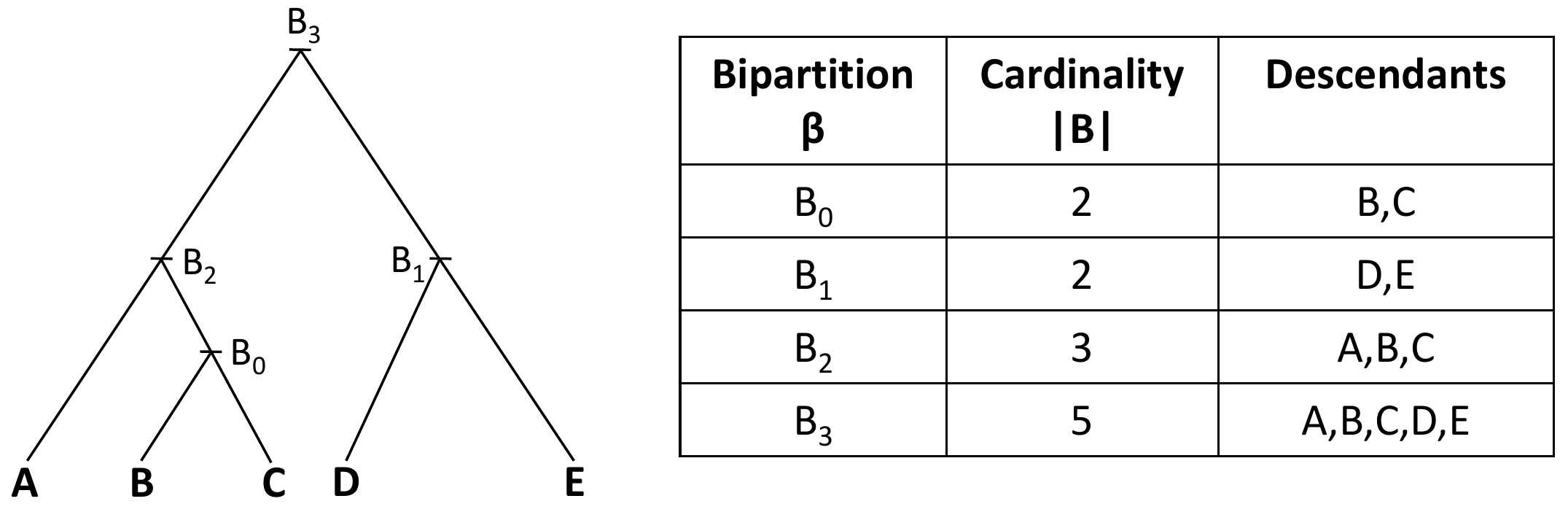}\caption{\label{fig:Original-tree-and}Original tree and its non-trivial and
root bipartitions for tree reconstruction example}
\end{figure}

\begin{figure}[tbh]
\includegraphics[scale=0.36]{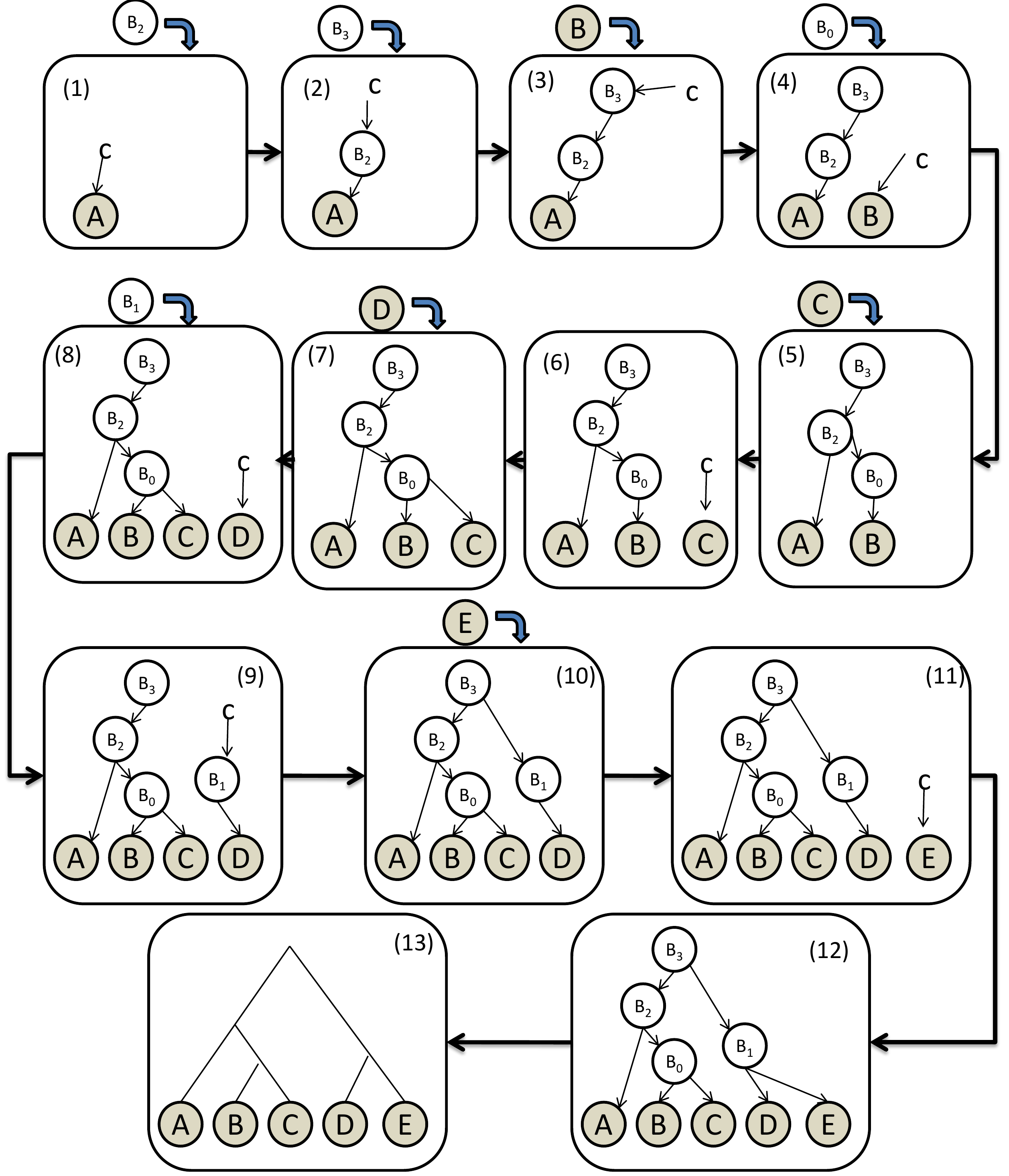}\caption{\label{fig:Tree-reconstruction-steps}Tree reconstruction steps in
EvoZip. Shaded circles represent leaf nodes and unshaded circles represent
internal nodes. Pointer $c$ points to the current node whose parent
is to be searched.}

\end{figure}

Figure \ref{fig:Original-tree-and} shows a sample five-taxon phylogenetic
tree along with its non-trivial and root bipartitions. Figure \ref{fig:Tree-reconstruction-steps}
demonstrates the working of EvoZip (Algorithm \ref{alg:Tree-reconstruction-algorithm})
by reconstructing the sample tree of Fig. \ref{fig:Original-tree-and}
using its bipartitions. The shaded and unshaded circles in Fig. \ref{fig:Tree-reconstruction-steps}
represent the leaf nodes and internal nodes respectively. The table
in Fig. \ref{fig:Original-tree-and} shows the set of all bipartitions,
represented by $\beta$, the cardinality of each bipartition, and
the leaf descendants of each bipartition.%
\begin{comment}
For instance, the bipartition $B_{2}$ has three leaf descendants,
namely $A,$$B,$ and $C$, so, its cardinality is three. Every leaf
in a phylogenetic tree is eventually a descendant of the root, hence,
the root bipartition in a tree has the highest cardinality. Thus,
the root bipartition, represented as $B_{3}$ in Fig. \ref{fig:Original-tree-and},
is always the last element of $\beta$.
\end{comment}
{} Once EvoZip has sorted the set $\beta$, it creates an empty tree
$T$ to start with. Let $L=\{A,B,C,D,E\}$ denote the set of all taxa
of the tree shown in Fig. \ref{fig:Original-tree-and}. EvoZip picks
a single taxon $A$ from $L$ and creates a leaf node corresponding
to $A$ in $T$. A pointer $c$ points towards $A$. In general, the
pointer $c$ always points to the current node whose parent is being
searched for in the tree being constructed. Having created the leaf
node $A$ (see Fig. \ref{fig:Tree-reconstruction-steps}.1), EvoZip
searches for the first bipartition $B$ in the sorted set $\beta$
containing leaf $A$, which happens to be $B_{2}$ (see Fig. \ref{fig:Original-tree-and}).
So, it attaches $A$ to $B_{2}$ as its child and the pointer $c$
now points to $B_{2}$ (Fig. \ref{fig:Tree-reconstruction-steps}.2).
Now, it searches for the next bipartition containing leaf $A$, which
is $B_{3}$, so, $B_{2}$ is attached to $B_{3}$ as a child (Fig.
\ref{fig:Tree-reconstruction-steps}.3). This completes the reconstruction
of the path from leaf $A$ to the root since no other bipartition
in $\beta$ contains $A$. Now, EvoZip takes the next taxa $B$ from
$L$ and makes $c$ point to it to indicate that $B$ is the next
node that has to be attached to its parent (Fig. \ref{fig:Tree-reconstruction-steps}.4).
The first bipartition containing $B$ is $B_{0}$, which is not yet
present in $T$, so, a new internal node $B_{0}$ is created and $B$
is attached to $B_{0}$ as a child. $c$ now points to the parent
of $B$, i.e., $B_{0}$ (Fig. \ref{fig:Tree-reconstruction-steps}.5).
The next bipartition containing $B$ is $B_{2}$. The internal node
$B_{2}$ already exists in $T$, i.e., the sub path from $B_{2}$
to the root overlaps with an already constructed leaf-to-root path.
Hence, $B_{0}$ is now attached to $B_{2}$ as its child and the reconstruction
of the path from leaf $B$ to the root is complete. This continues
until paths from all the leaves to the root node have been reconstructed
as shown in Fig. \ref{fig:Tree-reconstruction-steps}.

\begin{comment}
, and the next taxa $C$ is taken from $L$. A leaf node corresponding
to $C$ is added to $T$ and $c$ now points to the newly added leaf
$C$ (Fig. \ref{fig:Tree-reconstruction-steps}.6). Again, EvoZip
searches for bipartitions having $C$ as descendant. $B_{0}$ is the
first such bipartition. Since $B_{0}$ has already been added to $T$,
$C$ is attached to $B_{0}$ (Fig. \ref{fig:Tree-reconstruction-steps}.7)
and next taxa $D$ is taken from $L$. A new leaf corresponding to
$D$ is created in $T$ (Fig. \ref{fig:Tree-reconstruction-steps}.8).
The first bipartition having $D$ as its descendant is $B_{1}$ that
does not exists in $T$ yet, so, a new internal node $B_{1}$ is created
in $T$ and $D$ is attached to $B_{1}$ as its child, and $c$ is
made to point $B_{1}$(Fig. \ref{fig:Tree-reconstruction-steps}.9).
$B_{3}$ is the next bipartition in $\beta$ having $D$ as a descendant.
$B_{3}$ is already present in $T$, so, $B_{1}$ is attached to $B_{3}$
as its child (Fig. \ref{fig:Tree-reconstruction-steps}.10). Next
taxon $E$ is taken from $L$ and a new leaf corresponding to $E$
is created in $T$ (Fig. \ref{fig:Tree-reconstruction-steps}.11).
The first bipartition having $E$ as a descendant is $B_{1}$, so,
$E$ is attached to $B_{1}$ as a child (Fig. \ref{fig:Tree-reconstruction-steps}.12).
All the taxa have been included in $T$ and reconstruction of $T$
is now complete (Fig. \ref{fig:Tree-reconstruction-steps}.13).
\end{comment}

\begin{algorithm}[tbh]
\begin{algorithmic}[1]
\STATE $\beta \gets$ set of non-trivial bipartitions $\cup$ root bipartition \label{reconst.1} 
\STATE Sort $\beta$ on the basis of cardinality in the ascending order. Let $\beta=\{B_0,B_1,...,B_k\}$ represent the sorted set, where $B_k$ is the root bipartition. \label{reconst.2} 
\FOR{each taxon $l$}\label{reconst.forStart} 
	\STATE Create a leaf node corresponding to $l$. Let $c$ point to the node.
	\STATE Find the first bipartition $B\in\beta$ that contains $l$
	\WHILE{$B$ exists}
		\IF{a node exists for $B$}\label{reconst.ifStart} 
			\STATE Make $c$ child of that node
			\STATE \textbf{break}\label{reconst.ifStop} 
		\ELSE \label{reconst.elseStart} 
			\STATE Create a node for $B$
			\STATE Make $c$ child of that node
			\STATE Let $c$ point to its parent
			\STATE Find the next bipartition $B\in\beta$ that contains $l$
		\ENDIF \label{reconst.elseEnd} 
	\ENDWHILE
\ENDFOR \label{reconst.forEnd}
\end{algorithmic}\caption{\label{alg:Tree-reconstruction-algorithm}Tree reconstruction algorithm}
\end{algorithm}

We were unable to determine the reconstruction algorithm used by TreeZip,
hence, we are unable to comment on the extent to which the use of
Gusfield's algorithm helps EvoZip improve upon TreeZip. However, we
would like to point out that, to the best of our knowledge, Gusfield's
algorithm \cite{gusfield1991efficient} is the fastest approach for
reconstructing a phylogenetic tree from its constituent bipartitions.

\subsubsection{Additional Utilities}

Apart from the above improvements over TreeZip, EvoZip also provides
the following additional utility functionalities:
\begin{itemize}
\item \textbf{Internal node labels}: We have already discussed the importance
of internal node labels in phylogenetic trees and their presence in
Newick representation in Sec. \ref{sec:Introduction}. While internal
node labels are ignored by TreeZip, EvoZip retains them in the compressed
file.
\item \textbf{Preserving precision in branch lengths}: We observed that
TreeZip only supports six digits of precision in branch lengths. If
the precision exceeds this limit, it rounds-off the numbers to six
digit precision numbers. EvoZip supports full precision for branch
lengths.
\item \textbf{Check for double collision}: While populating the hash table,
the possibility of double collision is minuscule (less than one in
$10^{17}$ in our implementation). It is so small that we never encountered
a double collision in all our experiments. The same observation was
also supported in TreeZip \cite{matthews2010novel}. Still, there remains 
a theoretical possibility of double
collision, which,  as discussed in
Sec. \ref{sec:The-TreeZip-Algorithm}, can corrupt the compression output. 
EvoZip provides an option to
check for double collision and any other error that may have crept.
This is achieved by comparing the input trees with the decompressed
trees. Since phylogenetic trees are unordered trees, comparison of
two phylogenetic trees must ignore the order of nodes. For this we
employ a phylogeny-specific canonical form for evolutionary trees
used in our earlier work on frequent subtree mining \cite{deepak2014evominer}.
Canonicalization is achieved by assigning a virtual label to each
internal node. The virtual label of an internal node $v$ is the minimum
label among all the leaf descendants of $v$. The children of an internal
node are ordered from left to right based on the sequence in which
they are encountered in an inorder depth first traversal, the leftmost
child being encountered first. A tree is in canonical form if, for
every internal node, its children are ordered from left to right by
their virtual labels. EvoZip checks for double collision by first
canonicalizing the input and output (decompressed) trees, and then
comparing their canonical forms. A mismatch of canonical forms 
indicates a double collision.
\end{itemize}

\section{Experimental results and discussion \label{sec:Results-and-discussion}}

Here we report on our experiments to
compare EvoZip with TreeZip \cite{matthews2010novel} with respect
to the size of the compressed file and the time taken (compression time
and decompression time). 
We performed our experiments on a Linux machine having 3.6 GHz Intel
Core i7-4790 octa-core CPU with 8 GB RAM and 64 bit Ubuntu 16.04 LTS
operating system. We evaluated EvoZip on the phylogenetic dataset
used in a previous study by Pattengale et al. \cite{pattengale2010uncovering}.

The test data consists of 17 sets of bootstrapped trees, each having around 10000
trees, constructed from single-gene and multi-gene DNA sequences,
with 125-2,554 taxa. Their file sizes range from 9.66 MB to 229.19
MB (see Table \ref{tab:Dataset-properties}). In the subsequent discussion,
each dataset is referred to by the number of taxa and the number of
trees in the dataset separated by `/'. 

While comparing EvoZip with TreeZip, we 
also report, under the heading `EvoZip w/o Deflate', the corresponding values where EvoZip is used \emph{without} the Deflate algorithm. This has
been done to highlight the individual contributions of bipartition/support-list
encoding schemes and the Deflate algorithm in EvoZip. All sizes are
presented in KBs, while the time is in seconds.  Despite repeated
attempts, TreeZip was unable to produce results on datasets `1481/10000',
`1512/14485', and `2000/10015'. Hence, the corresponding entries in 
Figures \ref{fig:Comparision-of-file-size}-\ref{fig:percent-Decompression-time-100=000025}
and Tables \ref{tab:File-size-comparison-1}-\ref{tab:Decompression-time}
are absent.

\begin{table}[tbh]
\caption{\label{tab:Dataset-properties}Dataset properties}

%\begin{centering}
%(This dataset has previously been used by Pattengale et al. \cite{pattengale2010uncovering})\medskip{}
%\par\end{centering}
\begin{tabular}{|c|c|c|c|c|c|}
\hline 
\multirow{2}{*}{\textbf{Sl.}} & \multirow{2}{*}{\textbf{\#taxa}} & \multirow{2}{*}{\textbf{\#trees}} & \textbf{\#bipartitions} & \textbf{\#bipartitions} & \textbf{FileSize}\tabularnewline
 &  &  & \textbf{(All)} & \textbf{(Unique)} & \textbf{(MB)}\tabularnewline
\hline 
\hline 
1 & 125 & 10000 & 1230125 & 321 & 8.88\tabularnewline
\hline 
2 & 150 & 10000 & 1480150 & 11487 & 18.57\tabularnewline
\hline 
3 & 218 & 10000 & 2160218 & 23495 & 26.58\tabularnewline
\hline 
4 & 354 & 10000 & 3520354 & 170752 & 53.32\tabularnewline
\hline 
5 & 404 & 10000 & 4020404 & 73839 & 33.62\tabularnewline
\hline 
6 & 500 & 10000 & 4980500 & 43131 & 51.87\tabularnewline
\hline 
7 & 628 & 10000 & 6260628 & 75785 & 60.22\tabularnewline
\hline 
8 & 714 & 10000 & 7120714 & 61995 & 60.22\tabularnewline
\hline 
9 & 994 & 10150 & 10069794 & 65492 & 105.70\tabularnewline
\hline 
10 & 1288 & 12800 & 16462088 & 179531 & 143.68\tabularnewline
\hline 
11 & 1481 & 10000 & 14791481 & 720201 & 130.65\tabularnewline
\hline 
12 & 1512 & 14485 & 21873862 & 591498 & 229.19\tabularnewline
\hline 
13 & 1604 & 10455 & 16750514 & 353834 & 175.72\tabularnewline
\hline 
14 & 1908 & 10458 & 19934856 & 441188 & 179.22\tabularnewline
\hline 
15 & 2000 & 10015 & 20011970 & 763655 & 142.19\tabularnewline
\hline 
16 & 2308 & 10074 & 23232952 & 159121 & 211.07\tabularnewline
\hline 
17 & 2554 & 10112 & 25808378 & 472989 & 186.30\tabularnewline
\hline 
\end{tabular}
\end{table}

\subsection{Compression Ratio}

\begin{figure}[tbh]
\centering{}\includegraphics[scale=0.3]{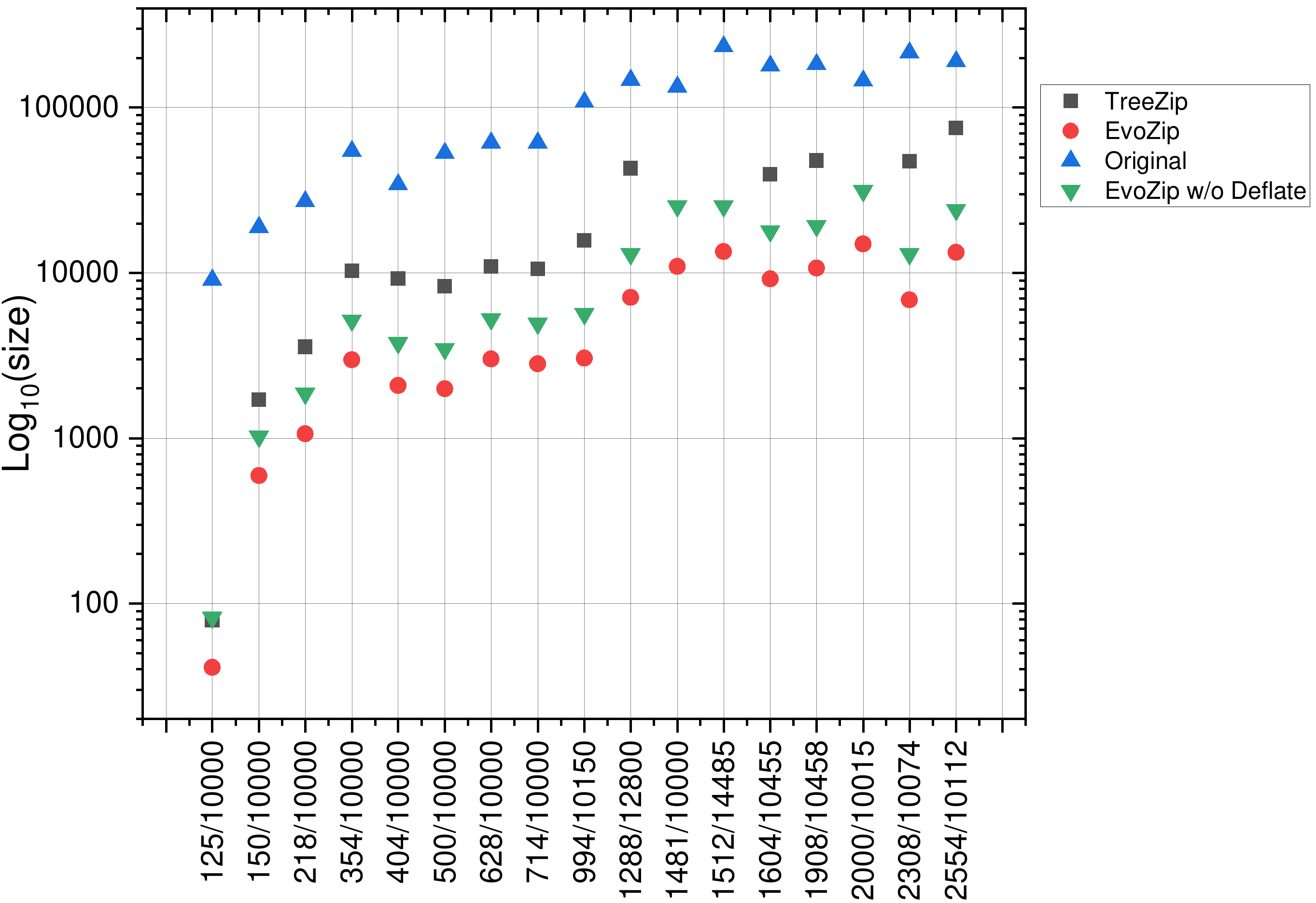}\caption{\label{fig:Comparision-of-file-size}Comparison of file size. The Y-axis shows size on a logarithmic scale.}
\end{figure}

\begin{figure}[tbh]
\centering{}\includegraphics[scale=0.3]{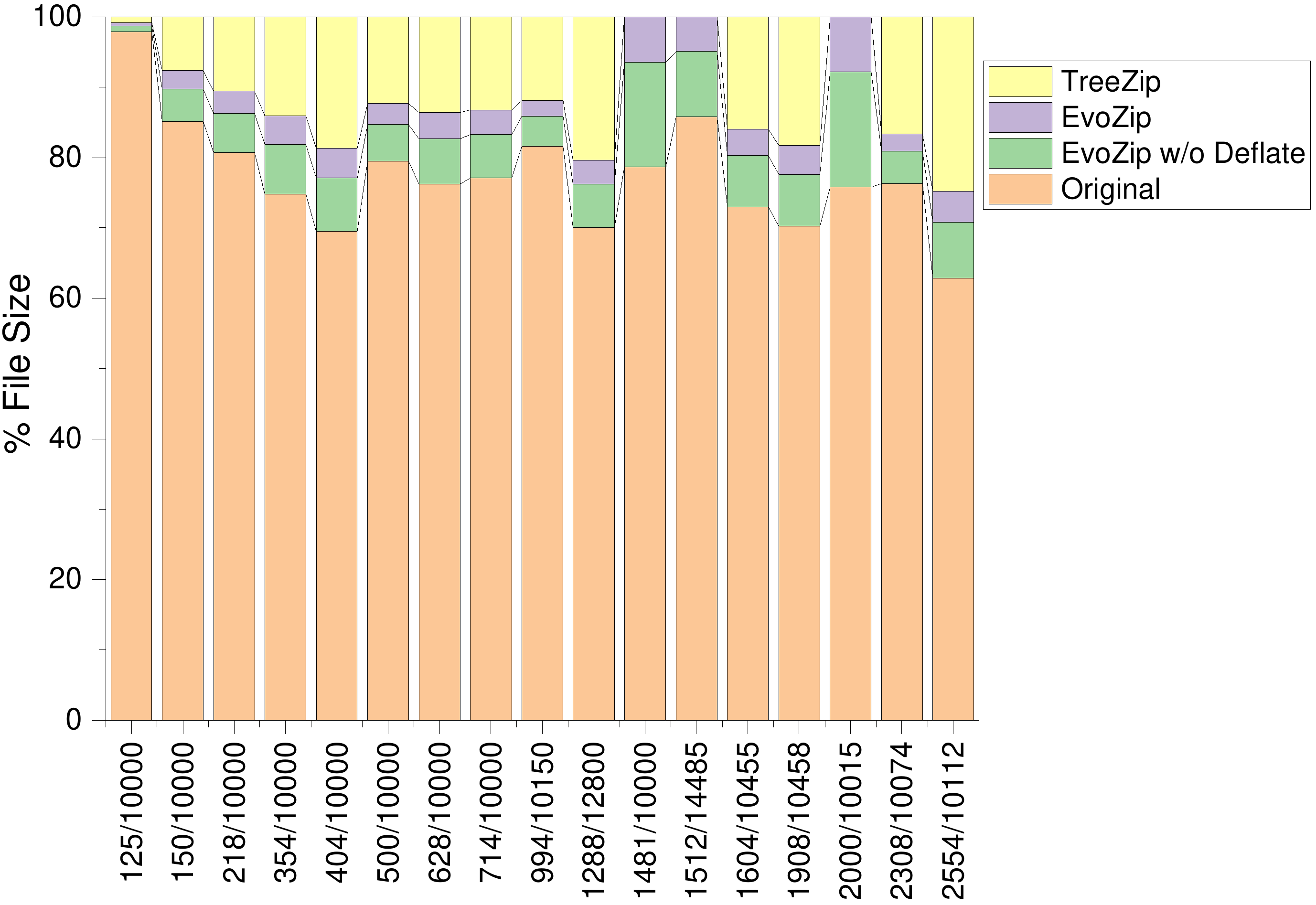}\caption{\label{fig:Percentage-comparision-of-file-size}Percentage comparison
of file size}
\end{figure}

Figure \ref{fig:Comparision-of-file-size} compares the original files
and the files compressed by TreeZip and EvoZip.  
Table \ref{tab:File-size-comparison-1} give the corresponding
values. On average, EvoZip produces 71.6\% better compression than TreeZip. Figure
\ref{fig:Percentage-comparision-of-file-size} shows the same comparison
on a relative scale.

\begin{table}[tbh]
\begin{centering}
\caption{\label{tab:File-size-comparison-1}File size comparison}
\par\end{centering}
\centering{}%
\begin{tabular}{|c|c|c|c|c|c|}
\hline 
\multirow{2}{*}{\textbf{Sl.}} & \multirow{2}{*}{\textbf{\#taxa/trees}} & \textbf{Size} & \multirow{2}{*}{\textbf{TreeZip}} & \multirow{2}{*}{\textbf{EvoZip}} & \textbf{EvoZip}\tabularnewline
 &  & \textbf{(KB)} &  &  & \textbf{w/o Deflate}\tabularnewline
\hline 
\hline 
1 & 125/10000 & 9093 & 79 & 41 & 82\tabularnewline
\hline 
2 & 150/10000 & 19014 & 1701 & 595 & 1021\tabularnewline
\hline 
3 & 218/10000 & 27217 & 3567 & 1066 & 1862\tabularnewline
\hline 
4 & 354/10000 & 54600 & 10281 & 2979 & 5163\tabularnewline
\hline 
5 & 404/10000 & 34424 & 9256 & 2079 & 3771\tabularnewline
\hline 
6 & 500/10000 & 53116 & 8253 & 1997 & 3480\tabularnewline
\hline 
7 & 628/10000 & 61670 & 10970 & 3016 & 5256\tabularnewline
\hline 
8 & 714/10000 & 61670 & 10574 & 2816 & 4939\tabularnewline
\hline 
9 & 994/10150 & 108241 & 15728 & 3057 & 5659\tabularnewline
\hline 
10 & 1288/12800 & 147125 & 42940 & 7092 & 12998\tabularnewline
\hline 
11 & 1481/10000 & 133790 & \textendash{} & 10924 & 25378\tabularnewline
\hline 
12 & 1512/14485 & 234689 & \textendash{} & 13456 & 25332\tabularnewline
\hline 
13 & 1604/10455 & 179941 & 39519 & 9191 & 17975\tabularnewline
\hline 
14 & 1908/10458 & 183526 & 47776 & 10732 & 19232\tabularnewline
\hline 
15 & 2000/10015 & 145599 & \textendash{} & 15025 & 31447\tabularnewline
\hline 
16 & 2308/10074 & 216139 & 47206 & 6867 & 13050\tabularnewline
\hline 
17 & 2554/10112 & 190776 & 75144 & 13353 & 24193\tabularnewline
\hline 
\end{tabular}
\end{table}

\begin{figure}[tbh]
\centering{}\includegraphics[scale=0.3]{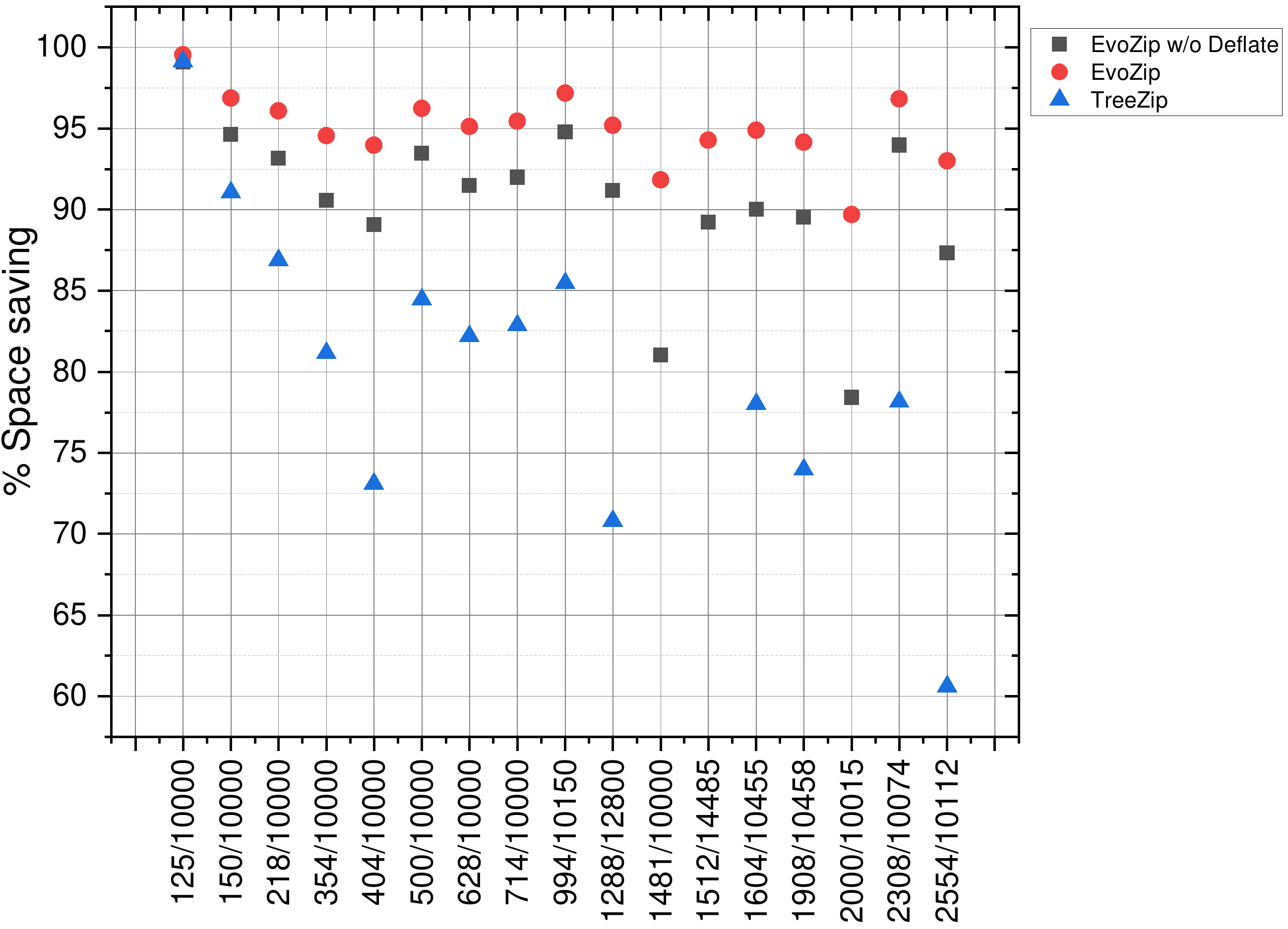}\caption{\label{fig:Space-savings}Percentage space savings}
\end{figure}

Figure \ref{fig:Space-savings} compares the percentage space savings
of EvoZip and TreeZip computed as follows
\[
\%\,\mathit{Space\,saving}=\left(1-\frac{1}{\mathit{Compression\,ratio}}\right)\times100
\]
where, the compression ratio is computed as
\[
\mathit{Compression\,ratio}=\frac{\mathit{Uncompressed\,size}}{\mathit{Compressed\,size}}
\]

Corresponding values are given in Table \ref{tab:Space-savings}.
On average, TreeZip achieves 80.57\% space saving while EvoZip achieves
94.99\%. Table \ref{tab:Compression-ratio} shows the corresponding
values for the compression ratio, based on the above
formula.

\begin{table}[tbh]
\begin{centering}
\caption{\label{tab:Space-savings}Comparison of percentage space saving}
\par\end{centering}
\centering{}%
\begin{tabular}{|c|c|c|c|c|}
\hline 
\multirow{2}{*}{\textbf{Sl.}} & \multirow{2}{*}{\textbf{\#taxa/trees}} & \multirow{2}{*}{\textbf{TreeZip}} & \multirow{2}{*}{\textbf{EvoZip}} & \textbf{EvoZip}\tabularnewline
 &  &  &  & \textbf{w/o Deflate}\tabularnewline
\hline 
\hline 
1 & 125/10000 & 99.1312 & 99.5491 & 99.09821\tabularnewline
\hline 
2 & 150/10000 & 91.05396 & 96.87073 & 94.63027\tabularnewline
\hline 
3 & 218/10000 & 86.89422 & 96.08333 & 93.15869\tabularnewline
\hline 
4 & 354/10000 & 81.17033 & 94.54396 & 90.54396\tabularnewline
\hline 
5 & 404/10000 & 73.11178 & 93.96061 & 89.04543\tabularnewline
\hline 
6 & 500/10000 & 84.46231 & 96.2403 & 93.4483\tabularnewline
\hline 
7 & 628/10000 & 82.21177 & 95.10945 & 91.47722\tabularnewline
\hline 
8 & 714/10000 & 82.8539 & 95.43376 & 91.99124\tabularnewline
\hline 
9 & 994/10150 & 85.46946 & 97.17575 & 94.77185\tabularnewline
\hline 
10 & 1288/12800 & 70.81393 & 95.17961 & 91.16534\tabularnewline
\hline 
11 & 1481/10000 & \textendash{} & 91.83497 & 81.03147\tabularnewline
\hline 
12 & 1512/14485 & \textendash{} & 94.26645 & 89.20614\tabularnewline
\hline 
13 & 1604/10455 & 78.0378 & 94.89221 & 90.01061\tabularnewline
\hline 
14 & 1908/10458 & 73.96772 & 94.15233 & 89.52083\tabularnewline
\hline 
15 & 2000/10015 & \textendash{} & 89.68056 & 78.40164\tabularnewline
\hline 
16 & 2308/10074 & 78.15943 & 96.82288 & 93.96222\tabularnewline
\hline 
17 & 2554/10112 & 60.6114 & 93.00069 & 87.31864\tabularnewline
\hline 
\end{tabular}
\end{table}

\begin{comment}
Figure \ref{fig:Compression-ratio} compares the compression ratio
of TreeZip with EvoZip. Corresponding values are given in Table \ref{tab:Compression-ratio}.
EvoZip achieves better compression ratios for all the datasets.
\end{comment}
\begin{comment}
\begin{figure}[tbh]
\centering{}\includegraphics[scale=0.3]{Images/CompressionRatio.pdf}\caption{\label{fig:Compression-ratio}Compression ratio}
\end{figure}
\end{comment}

\begin{table}[tbh]
\caption{\label{tab:Compression-ratio}Compression ratio}

\centering{}%
\begin{tabular}{|c|c|c|c|c|}
\hline 
\multirow{2}{*}{\textbf{Sl.}} & \multirow{2}{*}{\textbf{\#taxa/trees}} & \multirow{2}{*}{\textbf{TreeZip}} & \multirow{2}{*}{\textbf{EvoZip}} & \textbf{EvoZip}\tabularnewline
 &  &  &  & \textbf{w/o Deflate}\tabularnewline
\hline 
\hline 
1 & 125/10000 & 115.10 & 221.78 & 110.89\tabularnewline
\hline 
2 & 150/10000 & 11.18 & 31.96 & 18.62\tabularnewline
\hline 
3 & 218/10000 & 7.63 & 25.53 & 14.62\tabularnewline
\hline 
4 & 354/10000 & 5.31 & 18.33 & 10.58\tabularnewline
\hline 
5 & 404/10000 & 3.72 & 16.56 & 9.13\tabularnewline
\hline 
6 & 500/10000 & 6.44 & 26.60 & 15.26\tabularnewline
\hline 
7 & 628/10000 & 5.62 & 20.45 & 11.73\tabularnewline
\hline 
8 & 714/10000 & 5.83 & 21.90 & 12.49\tabularnewline
\hline 
9 & 994/10150 & 6.88 & 35.41 & 19.13\tabularnewline
\hline 
10 & 1288/12800 & 3.43 & 20.75 & 11.32\tabularnewline
\hline 
11 & 1481/10000 & \textendash{} & 12.25 & 5.27\tabularnewline
\hline 
12 & 1512/14485 & \textendash{} & 17.44 & 9.26\tabularnewline
\hline 
13 & 1604/10455 & 4.55 & 19.58 & 10.01\tabularnewline
\hline 
14 & 1908/10458 & 3.84 & 17.10 & 9.54\tabularnewline
\hline 
15 & 2000/10015 & \textendash{} & 9.69 & 4.63\tabularnewline
\hline 
16 & 2308/10074 & 4.58 & 31.48 & 16.56\tabularnewline
\hline 
17 & 2554/10112 & 2.54 & 14.29 & 7.89\tabularnewline
\hline 
\end{tabular}
\end{table}

\subsection{Compression/Decompression Time}

\begin{figure}[tbh]
\begin{centering}
\includegraphics[scale=0.3]{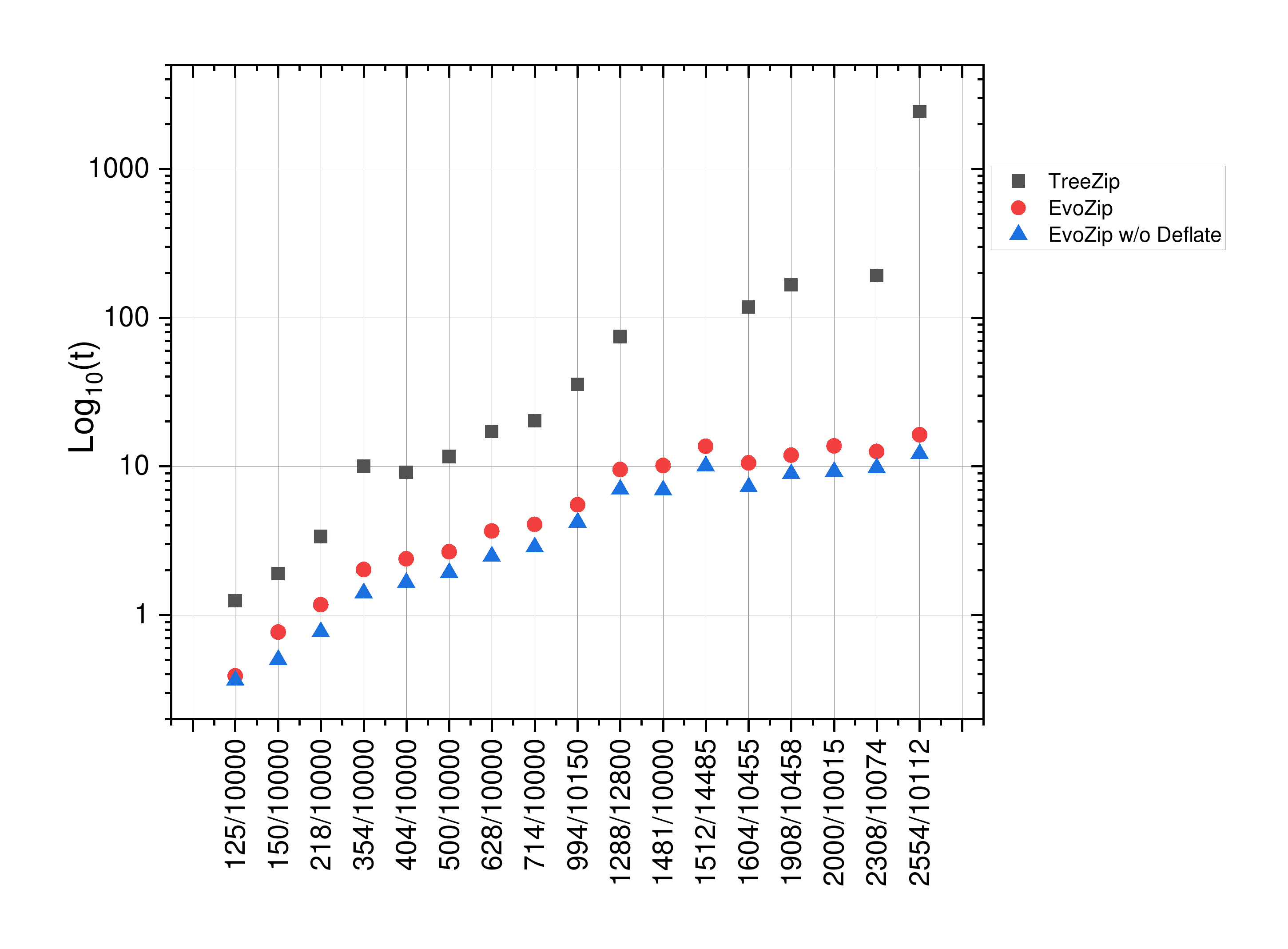}
\par\end{centering}
\caption{\label{fig:Comparision-of-compression-time}Comparison of compression
time. The time has been expressed on a logarithmic scale on Y-axis}
\end{figure}

\begin{figure}[tbh]
\centering{}\includegraphics[scale=0.3]{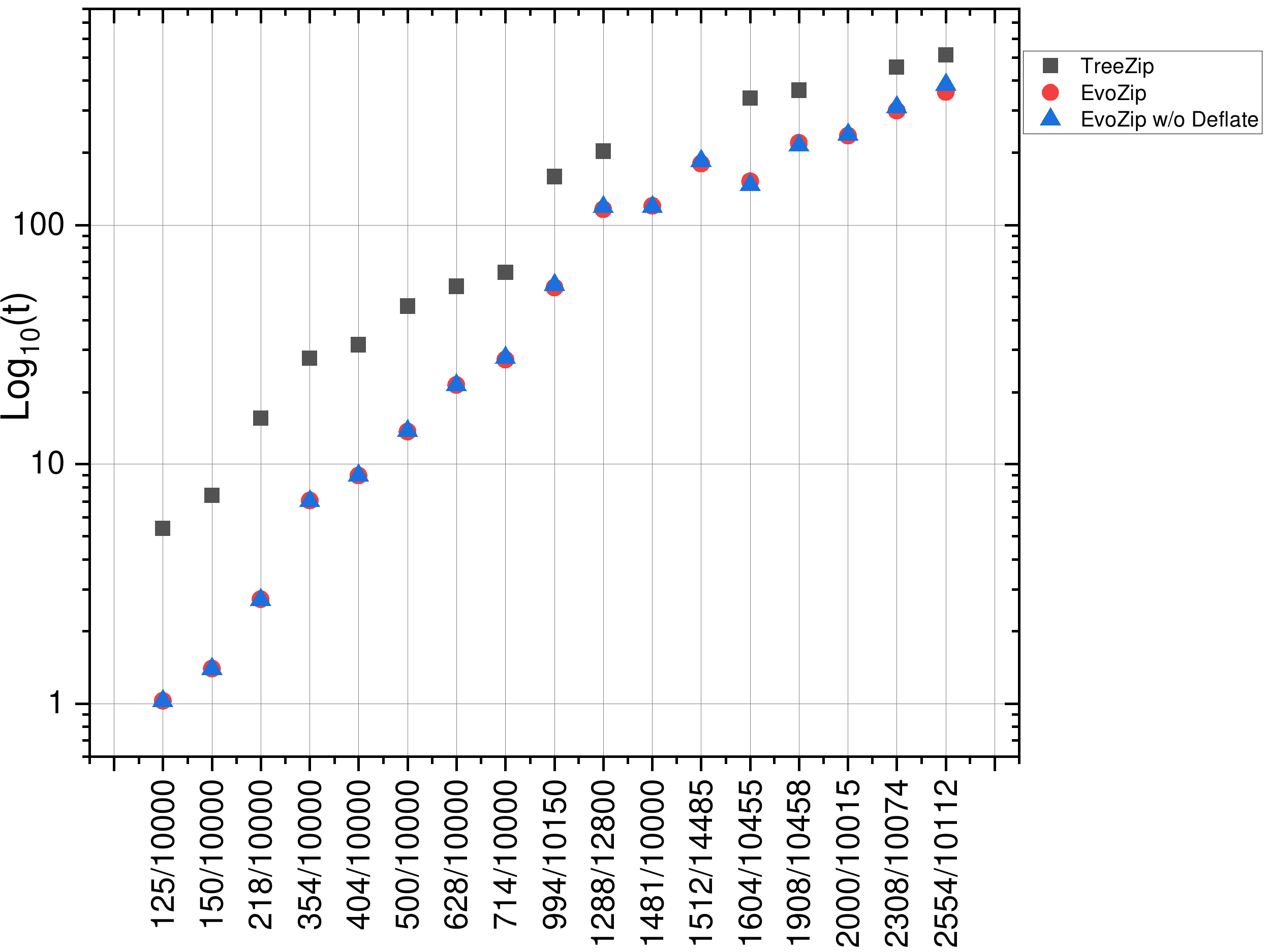}\caption{\label{fig:Comparision-of-decompression-time}Comparison of decompression
time. The time has been expressed on a logarithmic scale on Y-axis}
\end{figure}

\begin{figure}[tbh]
\centering{}\includegraphics[scale=0.3]{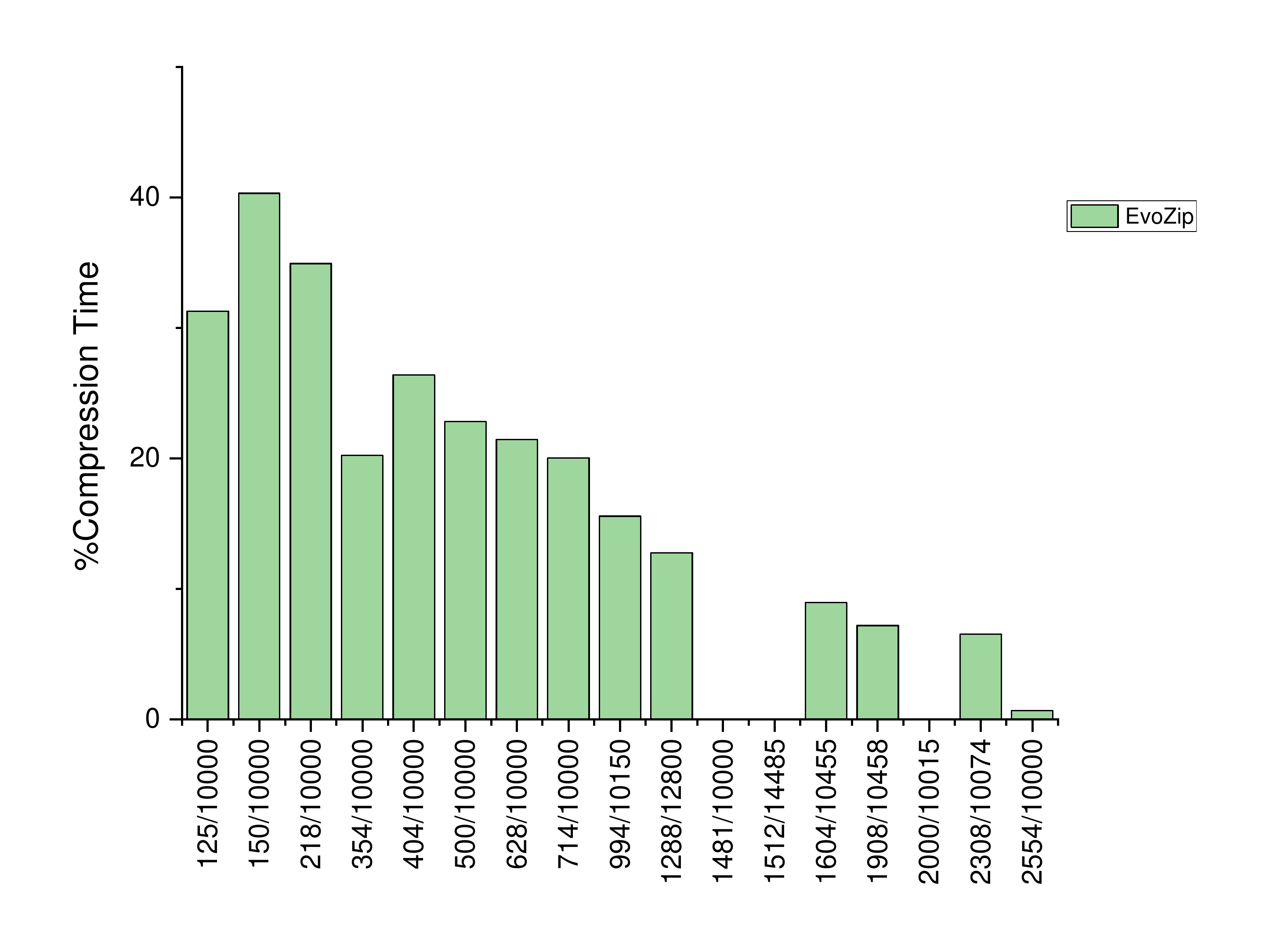}\caption{\label{fig:percent-Compression-time-100=000025}Percentage compression
time with respect to TreeZip. The above graph shows the compression
time taken by EvoZip as percentage of the corresponding time taken
by TreeZip.}
\end{figure}

\begin{figure}[tbh]
\centering{}\includegraphics[scale=0.3]{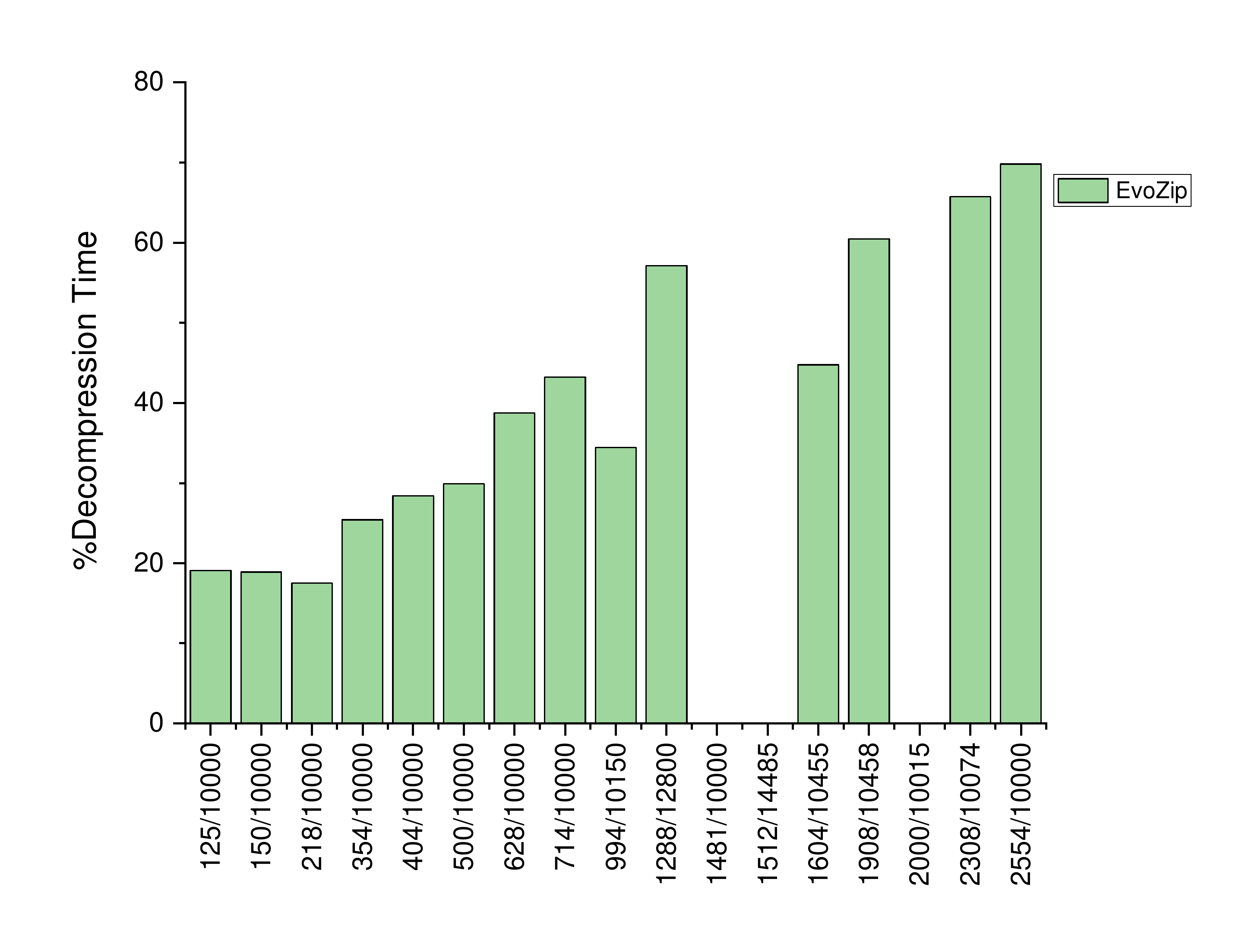}\caption{\label{fig:percent-Decompression-time-100=000025}Percentage decompression
time with respect to TreeZip. The above graph shows the compression
time taken by EvoZip as percentage of the corresponding time taken
by TreeZip.}
\end{figure}

%Next, the performances of TreeZip and EvoZip are compared with respect
%to time. 
Figures \ref{fig:Comparision-of-compression-time} and \ref{fig:Comparision-of-decompression-time}
compare compression and decompression times of EvoZip and TreeZip.
The corresponding values are given in Tables \ref{tab:Compression-time}
and \ref{tab:Decompression-time} respectively. EvoZip clearly beats
TreeZip for all sets of bootstrap trees,.  Note that EvoZip has better performance
gain for compression time than decompression time against TreeZip
(see Figures \ref{fig:percent-Compression-time-100=000025} and \ref{fig:percent-Decompression-time-100=000025}).
This is because during decompression the time to reconstruct the trees
from their constituent bipartitions dominates the overall decompression
time. For the smallest dataset (having 125 taxa and 10000 trees),
EvoZip (TreeZip) takes 0.391 (1.24944) seconds for compression, showing
about $3.2$ times improvement over TreeZip, and takes 1.028 (5.389062)
seconds for decompression, showing about $5.24$ times improvement
over TreeZip. Similarly, for the second largest dataset (having 2308
taxa and 10074 trees), EvoZip (TreeZip) takes 12.586 (192.147) seconds
for compression, a $15.27$-fold improvement, and takes
299.611 (455.999) seconds for decompression, a $1.52$-fold
improvement over TreeZip. The compressed file sizes of these
datasets using EvoZip (TreeZip) are 41KB (79KB) and 6867KB (47206KB)
respectively. We cannot comment on the improvement in
the compression time of the largest dataset (2554 taxa and
10112 trees) because, in repeated trials, TreeZip took an exceptionally long time for successful compression (2434 seconds), which is disproportionately high as compared to the time it took to compress other datasets (see Table  \ref{tab:Compression-time}). 
%\comm{Was the run aborted at 2434 secs.?}

\begin{comment}
\begin{figure}[tbh]
\caption{\label{fig:Time-comparison-charts}Time comparison}
\end{figure}
\end{comment}

\begin{table}[tbh]
\caption{\label{tab:Compression-time}Compression time}

\centering{}%
\begin{tabular}{|c|c|c|c|c|}
\hline 
\multirow{2}{*}{\textbf{Sl.}} & \multirow{2}{*}{\textbf{\#taxa/trees}} & \multirow{2}{*}{\textbf{TreeZip}} & \multirow{2}{*}{\textbf{EvoZip}} & \textbf{EvoZip}\tabularnewline
 &  &  &  & \textbf{w/o Deflate}\tabularnewline
\hline 
\hline 
1 & 125/10000 & 1.24944 & 0.391 & 0.366\tabularnewline
\hline 
2 & 150/10000 & 1.90709 & 0.769 & 0.504\tabularnewline
\hline 
3 & 218/10000 & 3.36997 & 1.177 & 0.772\tabularnewline
\hline 
4 & 354/10000 & 10.0219 & 2.028 & 1.415\tabularnewline
\hline 
5 & 404/10000 & 9.06688 & 2.393 & 1.658\tabularnewline
\hline 
6 & 500/10000 & 11.6548 & 2.661 & 1.938\tabularnewline
\hline 
7 & 628/10000 & 17.1729 & 3.683 & 2.491\tabularnewline
\hline 
8 & 714/10000 & 20.3026 & 4.068 & 2.873\tabularnewline
\hline 
9 & 994/10150 & 35.5257 & 5.534 & 4.227\tabularnewline
\hline 
10 & 1288/12800 & 74.666 & 9.522 & 7.015\tabularnewline
\hline 
11 & 1481/10000 & \textendash{} & 10.16 & 6.928\tabularnewline
\hline 
12 & 1512/14485 & \textendash{} & 13.686 & 10.095\tabularnewline
\hline 
13 & 1604/10455 & 117.834 & 10.562 & 7.253\tabularnewline
\hline 
14 & 1908/10458 & 166.166 & 11.928 & 8.916\tabularnewline
\hline 
15 & 2000/10015 & \textendash{} & 13.754 & 9.266\tabularnewline
\hline 
16 & 2308/10074 & 192.147 & 12.586 & 9.822\tabularnewline
\hline 
17 & 2554/10112 & 2434 & 16.324 & 12.183\tabularnewline
\hline 
\end{tabular}
\end{table}

\begin{table}[tbh]

\caption{\label{tab:Decompression-time}Decompression time}

\begin{centering}
\begin{tabular}{|c|c|c|c|c|}
\hline 
\multirow{2}{*}{\textbf{Sl.}} & \multirow{2}{*}{\textbf{\#taxa/trees}} & \multirow{2}{*}{\textbf{TreeZip}} & \multirow{2}{*}{\textbf{EvoZip}} & \textbf{EvoZip}\tabularnewline
 &  &  &  & \textbf{w/o Deflate}\tabularnewline
\hline 
\hline 
1 & 125/10000 & 5.389062 & 1.028 & 1.027\tabularnewline
\hline 
2 & 150/10000 & 7.398401 & 1.397 & 1.398\tabularnewline
\hline 
3 & 218/10000 & 15.60293 & 2.731 & 2.723\tabularnewline
\hline 
4 & 354/10000 & 27.739307 & 7.054 & 7.023\tabularnewline
\hline 
5 & 404/10000 & 31.605248 & 8.978 & 8.982\tabularnewline
\hline 
6 & 500/10000 & 45.778981 & 13.689 & 13.82\tabularnewline
\hline 
7 & 628/10000 & 55.28115 & 21.419 & 21.458\tabularnewline
\hline 
8 & 714/10000 & 63.189929 & 27.33 & 27.949\tabularnewline
\hline 
9 & 994/10150 & 158.80178 & 54.651 & 56.291\tabularnewline
\hline 
10 & 1288/12800 & 203.04748 & 115.926 & 119.208\tabularnewline
\hline 
11 & 1481/10000 & \textendash{} & 119.959 & 118.742\tabularnewline
\hline 
12 & 1512/14485 & \textendash{} & 180.147 & 185.478\tabularnewline
\hline 
13 & 1604/10455 & 338.93075 & 151.661 & 147.265\tabularnewline
\hline 
14 & 1908/10458 & 363.87119 & 220.053 & 214.591\tabularnewline
\hline 
15 & 2000/10015 & \textendash{} & 235.599 & 238.557\tabularnewline
\hline 
16 & 2308/10074 & 455.99994 & 299.611 & 310.969\tabularnewline
\hline 
17 & 2554/10112 & 513.3841 & 358.417 & 386.321\tabularnewline
\hline 
\end{tabular}
\par\end{centering}
\end{table}

%All the above results conclusively demonstrate the significant improvement
%of EvoZip over TreeZip with respect to size and time.

\section{Conclusions and future work \label{sec:Conclusions-and-future}}

Our results show that the proposed algorithm EvoZip improves significantly upon  TreeZip,
the state-of-the-art algorithm for phylogenetic compression, 
with respect to both compression ratio and comprssion time. The improvement is the
result of (1) improved bipartition and support list encoding schemes,
(2) the use of Deflate compression algorithm, and (3) the use of Gusfield's
\cite{gusfield1991efficient} efficient algorithm for phylogenetic
tree reconstruction. EvoZip also provides some additional utilities
not supported by TreeZip. The current implementation of EvoZip will be made available online at the time of publication for use by scientific community . 

Currently, EvoZip supports only collections of phylogenetic trees 
with identical taxon sets. 
We plan to add support for heterogeneous
tree collections where each tree can have a different sets of taxa. Mining frequent subtrees \cite{deepak2014evominer,ramu2012scalable} in a collection of heterogeneous trees can also be implemented. 
Further, a feature to construct the majority rule consensus tree from
the compressed data, without reconstructing all the input trees, can
also be implemented. This will reduce the processing time when only
a consensus tree on a subset of taxa needs to be assessed rather than
the extraction of the entire collection of trees (e.g., as in the
case of STBase \cite{stBase2015}). An option for stability analysis of phylogenetic trees \cite{sheikh2012stability} can also be incorporated when sequence data is also available along with the trees in order to weed-out any unstable tree from a large tree collection. 

\section*{Acknowlegments}

This work was partially funded by Early Career Research grant from
the Department of Science \& Technology - Science \& Engineering Research
Board (DST-SERB), project file no. ECR/2016/001077, Government of
India.

Akshay Deepak has been awarded Young Faculty Research Fellowship (YFRF)
of Visvesvaraya PhD Programme of Ministry of Electronics \& Information
Technology, MeitY, Government of India. In this regard, he would like
to acknowledge that this publication is an outcome of the R\&D work
undertaken in the project under the Visvesvaraya PhD Scheme of Ministry
of Electronics \& Information Technology, Government of India, being
implemented by Digital India Corporation (formerly Media Lab Asia).

\bibliographystyle{ieeetr}
\bibliography{PhyloCompress}

\begin{IEEEbiography}[{\includegraphics[scale=0.1]{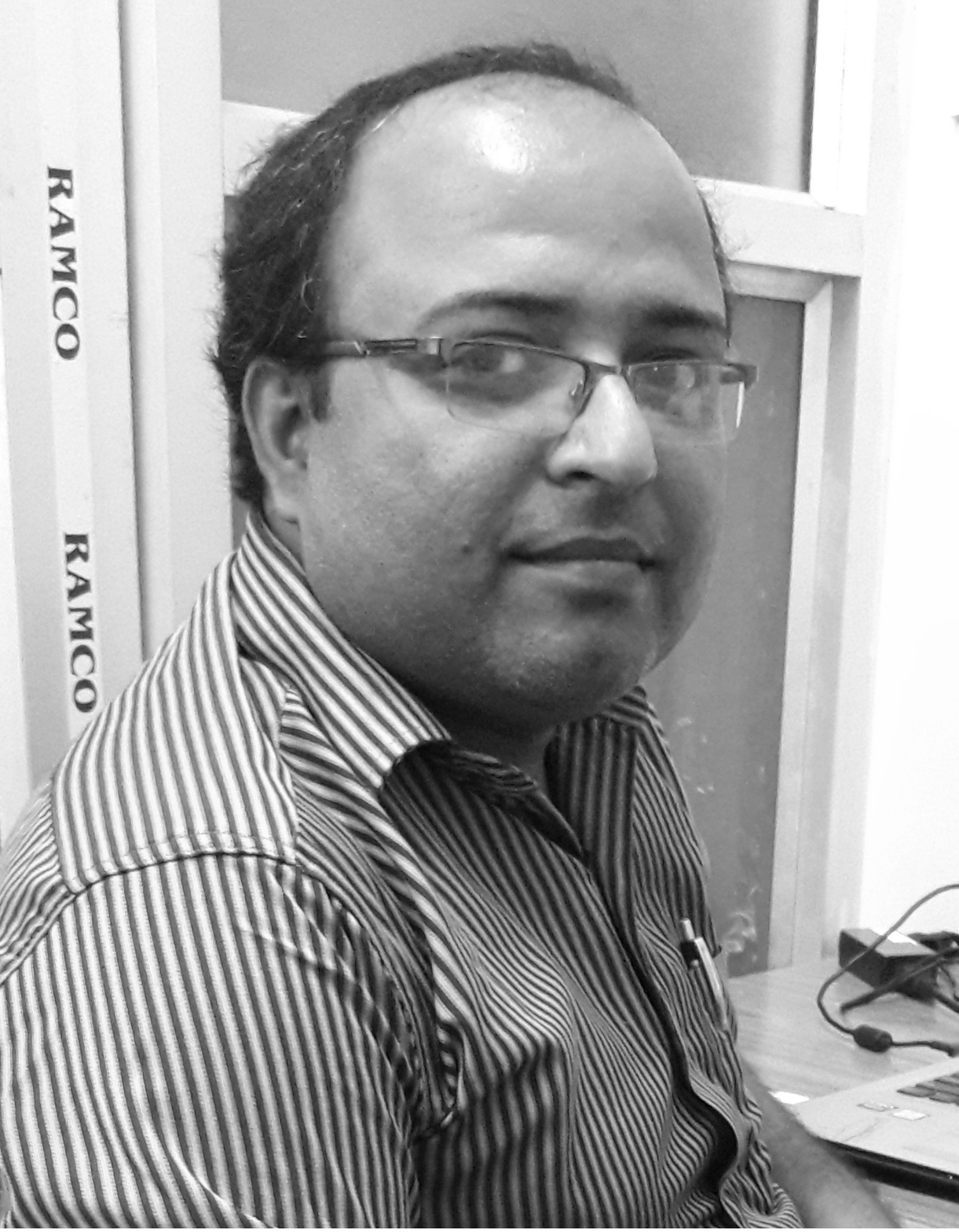}}]{Balanand Jha}
  is currently working as a senior research fellow in the Department
of Computer Science \& Engineering, National Institute of Technology
Patna, India. His research areas include computational phylogenetics,
ontologies, and computational biology.
\end{IEEEbiography}

\begin{IEEEbiography}[{\includegraphics[scale=0.04]{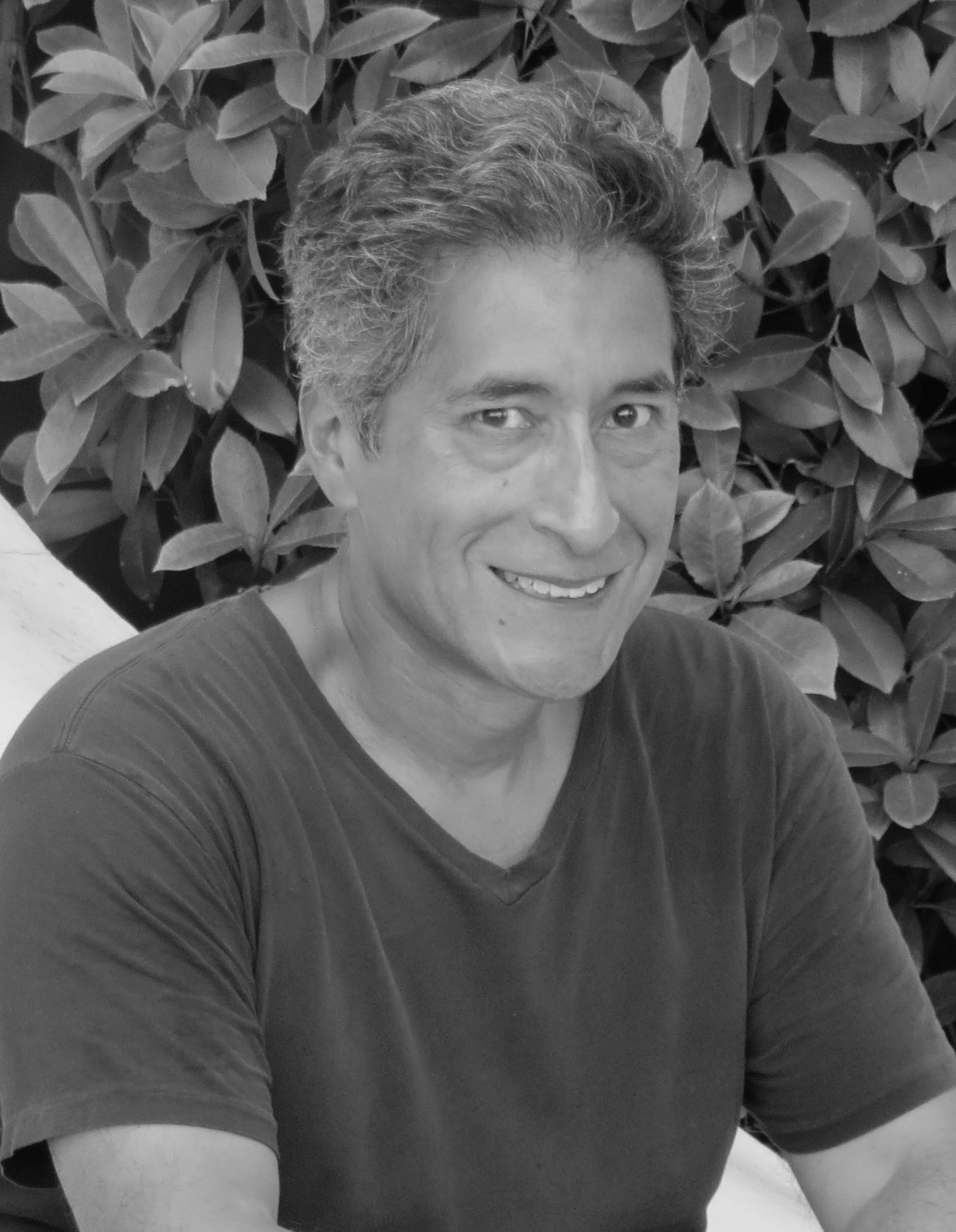}}]{David Fernández-Baca}
  is a professor in the Department of Computer Science, Iowa State
University, USA. His research interest includes combinatorial algorithms,
phylogenetic tree construction, computational biology, and sensitivity
analysis of optimization problems. \url{http://web.cs.iastate.edu/~fernande/}.
\end{IEEEbiography}

\begin{IEEEbiography}[{\includegraphics[scale=0.1]{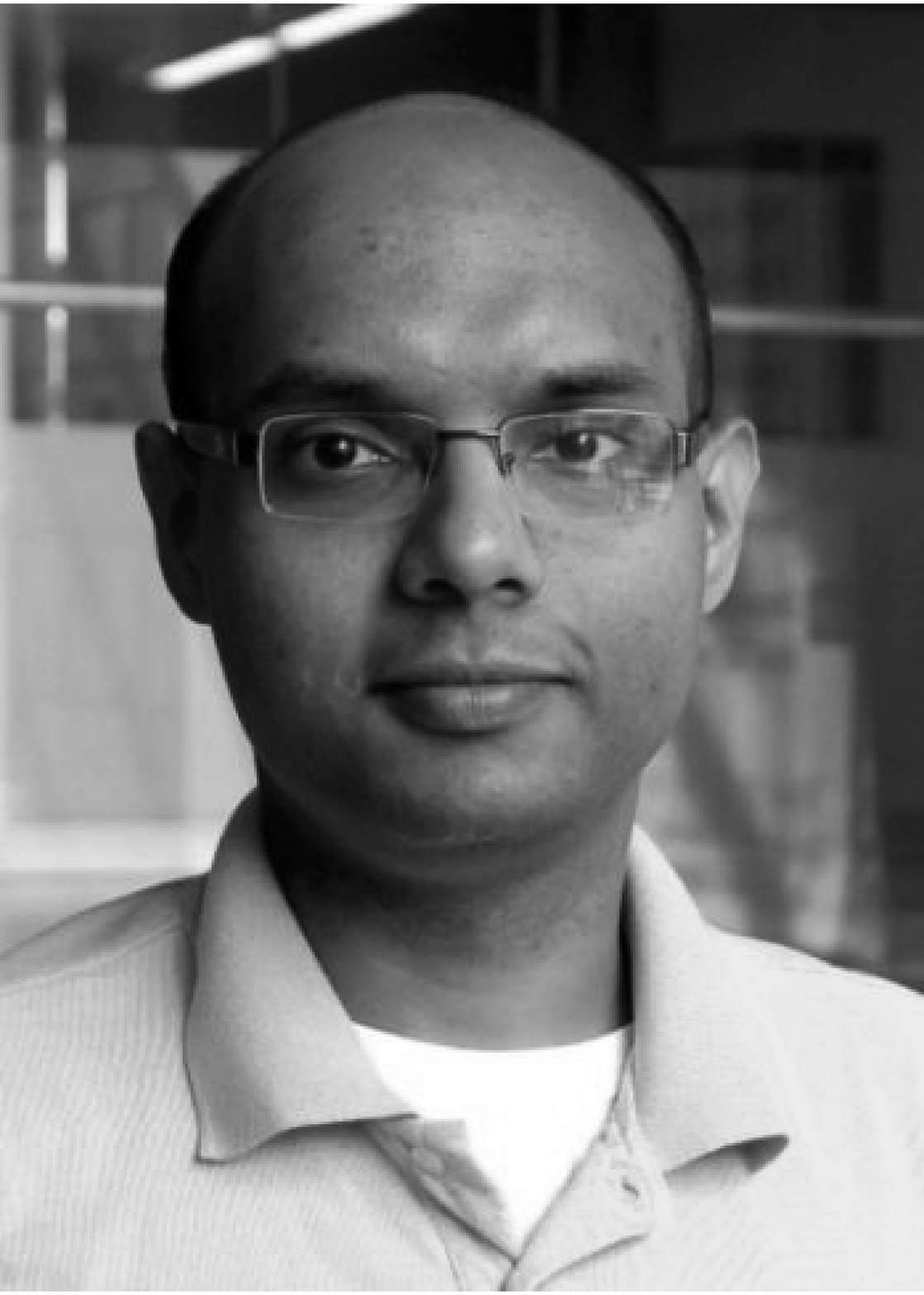}}]{Akshay Deepak}
  is an assistant professor in the Department of Computer Science
and Engineering, National Institute of Technology Patna, India. He
received his MS and PhD degrees in Computer Science from Iowa State
University, USA. His research interest include data mining, machine
learning, and computational biology. \url{http://www.nitp.ac.in/php/faculty_profile.php?id=akshayd@nitp.ac.in}.
\end{IEEEbiography}

\begin{IEEEbiography}[{\includegraphics[scale=0.1]{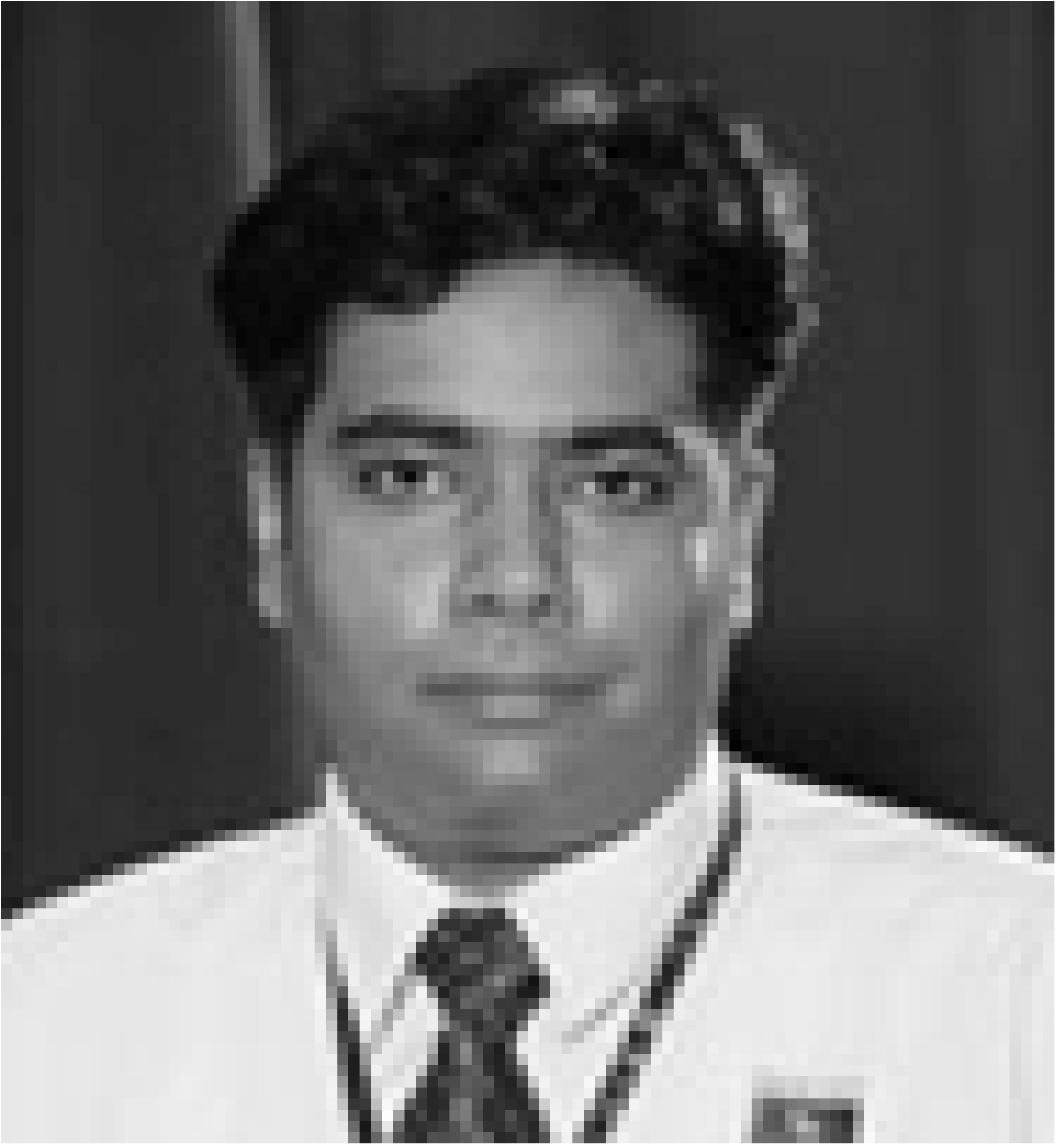}}]{Kumar Abhishek}
  is working as an Assistant Professor, Department of Computer Science
and Engineering, National Institute of Technology Patna, India. His
area of interest lies in RDF, Semantic Web, Ontology, Semantic Sensor
Web, and Big data. \url{http://www.nitp.ac.in/php/faculty_profile.php?id=kumar.abhishek@nitp.ac.in}
\end{IEEEbiography}

\end{document}